\documentclass[15pt]{article}
\usepackage{amssymb}
\usepackage{epsfig}

\newcommand{\BE}{\begin{equation}}
\newcommand{\EE}{\end{equation}}

\newcommand{\del}{{\mbox{\boldmath $\nabla$}}}
\newcommand{\half}{{\scriptstyle{\frac{1}{2}}}}
\newcommand{\vol}{{\sf V}}
\newcommand{\num}{{\sf N}}

\begin{document}

\vspace*{1mm}
\begin{center}

\vskip 1 pt

  {\Large{\bf {Quantum Non-Locality and the
CMB: what Experiments say}}}

\end{center}

\begin{center}
\vspace*{14mm} {\bf  Maurizio Consoli$^{(a)}$, Alessandro Pluchino$^{(b,a)}$, Paola Zizzi$^{(c)}$}
\vspace*{4mm}\\
{a) Istituto Nazionale di Fisica Nucleare, Sezione di Catania, Italy ~~~~~~~~~\\
b) Dipartimento di Fisica e Astronomia dell'Universit\`a di Catania,
Italy~\\
c) Department of Brain and Behavioural Sciences, Univ. of Pavia, Italy}
\end{center}

\par\noindent ``Non-Locality is most naturally incorporated into a
theory in which there is a special frame of reference. One possible
candidate for this special frame of reference is the one in which
the Cosmic Microwave Background (CMB) is isotropic. However, other
than the fact that a realistic interpretation of quantum mechanics
requires a preferred frame and the CMB provides us with one, there
is no readily apparent reason why the two should be linked'' (L.
Hardy). Starting from this remark we first argue that, given the
present view of the vacuum, the basic tenets of Quantum Field Theory
cannot guarantee that Einstein Special Relativity, with no preferred
frame, is the {\it physically realized} version of relativity. Then,
to try to understand the nature of the hypothetical preferred
$\Sigma-$frame, we consider the so called ether-drift experiments,
those precise optical measurements that try to detect in laboratory
a small angular dependence of the two-way velocity of light and then
to correlate this angular dependence with the direct CMB
observations with satellites in space. By considering all
experiments performed so far, from Michelson-Morley to the present
experiments with optical resonators, and analyzing the small
observed residuals in a modern theoretical framework, the long
sought $\Sigma-$ frame tight to the CMB naturally emerges. Finally,
if quantum non-locality reflects some effect propagating at vastly
superluminal speed $v_{QI} \to \infty $, its ultimate origin could
be hidden somewhere in the infinitely large speed $c_s \to \infty$ of the vacuum
density fluctuations.



\section{Introduction}

In spite of its extraordinary success in the description of
experiments, many conceptual aspects of Quantum Mechanics are still
puzzling \footnote{According to Weinberg ``It is a
bad sign that those physicists today who are most comfortable with
quantum mechanics do not agree with one another about what it all
means''\cite{weinberg}; or, according to  Blanchard, Fr\"ohlich and
Schubnel, ``Given that quantum mechanics was discovered ninety years
ago, the present rather low level of understanding of its deeper
meaning may be seen to represent some kind of intellectual
scandal''\cite{BFS}.}. In this paper, we will focus on a particular
aspect which is perhaps the most controversial: the violation of
Einstein locality and the conflict with (Einstein) relativity. This
has been the subject of a long debate which started with the seminal
paper by Einstein-Podolski-Rosen (EPR) \cite{EPR}, it was
substantially influenced by the work of Bell \cite{Bell}, and
continues unabated until today. To have an idea, among the various
implications, one may arrive to conclude that ``a free choice made
by an experimenter in one space-time region can influence a second
region that is space-like separated from the first'' \cite{Stapp}
(but see \cite{Shimony}).

For completeness, we observe that the problem dates back to the very
early days of Quantum Mechanics, even before EPR. Indeed, the basic
issue is already found in Heisenberg's 1929 Chicago Lectures: `` We
imagine a photon represented by a wave packet... By reflection at a
semi-transparent mirror, it is possible to decompose into a
reflected and a transmitted packet...After a sufficient time the two
parts will be separated by any distance desired; now if by
experiment the photon is found, say, in the reflected part of the
packet, then the probability of finding the photon in the other part
of the packet immediately becomes zero. The experiment at the
position of the reflected packet thus exerts a kind of action
(reduction of the wave packet) at the distant point and one sees
that this action is propagated with a velocity greater than that of
light''.

Then, Heisenberg adds immediately the following remark: ``However,
it is also obvious that this kind of action can never be utilized
for the transmission of signals so that it is not in conflict with
the postulates of relativity''. But, one may ask, if there were really something
which propagates nearly instantaneously, could such extraordinary
thing be so easily dismissed? Namely, could we ignore this
`something' just because it cannot be efficiently controlled to send
`messages' \cite{Bricmont0}? \footnote{``The impossibility of
sending messages is sometimes taken to mean that there is nothing
nonlocal going on. But nonlocality refers here to causal
interactions as described (in principle) by physical theories.
Messages are far more anthropocentric than that, and require that
humans be able to control these interactions in order to
communicate. As remarked by Maudlin \cite{Maudlin}, the Big Bang and
earthquakes cannot be used to send messages, but they have causal
effects nevertheless'' \cite{Bricmont0}.}. After all, this explains
why Dirac, more than forty years later, was still concluding that
``The only theory which we can formulate at the present is a
non-local one, and of course one is not satisfied with such a
theory. I think one ought to say that the problem of reconciling
quantum theory and relativity is not solved'' \cite{Dirac}.

Reduced to its essential terms, this locality problem concerns the
time ordering of two events A and B along the world line of a
hypothetical effect propagating with speed $> c$. This
ordering would be different in different frames, because in some
frame $S'$ one could find $t'_A
> t'_B$ and in some other frame $S''$ the opposite $t''_B
> t''_A$. This causal paradox is the main reason why superluminal signals are
not believed to exist. But, for instance, one cannot exclude
superluminal sound, i.e. density fluctuations propagating with speed
$c_s>c$. In fact, ``it is an open question whether $c_s/c$ remains
less than unity when non-electromagnetic forces are taken into
account''\cite{Weinbergsound}. Thus superluminal sound cannot be
excluded but is confined to very dense media and thus considered irrelevant
for the vacuum. As we shall see at the end of Sect.4, however, the
physical vacuum is a peculiar medium and this conclusion may be too naive.

Therefore, in principle, one could also change perspective and try
to dispose of the causality paradox if there were a preferred
reference system $\Sigma$ where the
superluminal effect propagates isotropically, see e.g. \cite{undivided,Hardy,Caban,Sonego,Eberhard1,Eberhard2}.
More explicitly, in
our case at hand of Quantum Mechanics, if we look at the ``Quantum
Information'' as a transport phenomenon \cite{garisto} which
propagates in space with superluminal velocity $v_{QI}\gg c$,
Quantum Mechanics corresponds to the
$v_{QI}\to \infty$ limit. Equivalently, by comparing with
experiments, one can set a lower limit on $v_{QI}$.
Assuming that the preferred
$\Sigma-$frame coincides with the reference system where the
Cosmic Microwave Background (CMB) is exactly isotropic \footnote{The system where the CMB Kinematic Dipole \cite{yoon} vanishes describes a
motion of the solar system with average velocity $V_{\rm CMB}\sim
370$ km/s, right ascension $\alpha_{\rm CMB} \sim 168^o$ and
declination $\gamma_{\rm CMB}\sim -7^o$, approximately pointing
toward the constellation Leo.}, present
experimental determinations give vastly superluminal values whose lower limit
has now increased from the original $v_{QI}> 10^4 c$
\cite{scarani, cocciaro, salart, bancal} up  to the more recent determination
$v_{QI}> 10^6 c$ \cite{cocciaro2}.

A frequent objection to the idea of relativity with a preferred
frame is that, after all, Quantum Mechanics is not a fundamental
description of the world. One should instead start from a
fundamental Quantum Field Theory (QFT) which incorporates the
locality requirement. More precisely, there are violations of
(micro)causality in QFT which originate from the lack of sharp
localizability of relativistic quantum systems \cite{maiani}.
However, these violations are confined to such small scales to be
completely irrelevant for the problem that we are considering. On
this basis, some authors have concluded that Bell's proof of
non-locality is either wrong or can just be used to rule out a
particular class of hidden-variable theories\footnote{This reductive
interpretation of Bell's work is contested by Bricmont
\cite{Bricmont}. Spelling out precisely the meaning of Bell's
theorem, he is very explicit on this point: ``Bell's result,
combined with the EPR argument, is rather that there are nonlocal
physical effects (and not just correlations between distant events)
in Nature''.}.

Therefore, if we adopt the perspective of an underlying, fundamental
QFT, solving the problem of locality in Quantum Mechanics could be
reduced to find a particular, missing logical step which prevents to
deduce, from the basic tenets of QFT, that Einstein Special
Relativity, with no preferred frame, is the {\it physically realized}
version of relativity. This is the version which is always assumed
when computing S-matrix elements for microscopic processes. However,
what one is actually using is the machinery of Lorentz
transformations whose first, complete derivation dates back,
ironically, to Larmor and Lorentz who were assuming the existence of
a fundamental state of rest (the ether). Our point here is that
there is indeed a particular element which has been missed so far
and depends on the nature of the vacuum state. Most likely, this is
{\it not} Lorentz invariant due to the phenomenon of vacuum
condensation, i.e. due to the macroscopic occupation of the same
quantum state.

In the physically relevant case of the Standard Model, the
phenomenon of vacuum condensation can be summarized by saying that
``What we experience as empty space is nothing but the configuration
of the Higgs field that has the lowest possible energy. If we move
from field jargon to particle jargon, this means that empty space is
actually filled with Higgs particles. They have Bose condensed''
\cite{thooft} \footnote{The explicit translation from field jargon
to particle jargon, with the substantial equivalence between the
effective potential of quantum field theory and the energy density
of a dilute particle condensate, can be found for instance in
ref.\cite{mech}, see also the following Sect.2.}. Clearly, this type
of medium is not the ether of classical physics. However, it is also
different from the 'empty'  space-time of Special Relativity that
Einstein had in mind in 1905 \footnote{In connection with the idea
of ether, it should be better underlined that Einstein's original
point of view had been later reconsidered with the transition from
Special Relativity to General Relativity \cite{Kostro}. Most
probably, he realized that Riemannian geometry is also the natural
framework to describe the dynamics of elastic media, see e.g. A.
Sommerfeld, Mechanics of Deformable Bodies, Academic Press, New York
1950.}.

To our knowledge, the idea that the phenomenon of vacuum
condensation could produce `conceptual tensions' with the basic locality of both Special and General
Relativity, was first discussed by Chiao \cite{Chiao}: ``The
physical vacuum, an intrinsically nonlocal ground state of a
relativistic quantum field theory, which possesses certain
similarities to the ground state of a superconductor... This would
produce an unusual `quantum rigidity' of the system, associated with
what London called the `rigidity of the macroscopic wave
function'... The Meissner effect is closely analog to the Higgs
mechanism in which the physical vacuum also spontaneously breaks
local gauge invariance''\footnote{After these arguments, Chiao
immediately adds the usual remark about the impossibility of
information propagating at superluminal speed: ``Relativistic
causality forbids only the front velocity, i.e., the velocity of
discontinuities, which connects causes to their effects, from
exceeding the speed of light, but does not forbid a wave packet
group velocity from being superluminal''\cite{Chiao}.}. Therefore,
it is not inconceivable that the macroscopic occupation of the same
quantum state, say ${\bf k}=0$ in some reference system $\Sigma$,
can represent the origin of the sought preferred frame. In
particular, as we will discuss in Sect.2, imposing that only local, scalar operators
(as the Higgs field, or the gluon condensate, or the
chiral condensate...) acquire a non-zero vacuum expectation value
does {\it not} imply the much stronger requirement of an exact
Lorentz-invariant vacuum state.

Since our arguments in Sect.2 are rather formal and give no
information on the nature of the preferred frame, we will then look
for definite experimental indications. As anticipated, existing
lower limits on $v_{QI}$ have assumed that $\Sigma$ is tight to the
CMB. But, as remarked by Hardy \cite{Hardy}, there is no readily
apparent reason for this identification. Therefore, to find the link
we will consider the so called ether-drift experiments where, by
precise optical measurements, one tries i) to detect in laboratory a
small angular dependence $\frac{ \Delta\bar{c}_\theta} {c} \neq 0$
of the two-way velocity of light and then ii) to correlate this
angular dependence with the direct CMB observations with satellites
in space.

Of course, experimental evidence for both the undulatory and
corpuscular aspects of radiation has substantially modified the
consideration of an underlying ethereal medium, as support of the
electromagnetic waves, and its logical need for the physical theory.
Yet, by accepting the idea of a preferred frame, the final physical description could become qualitatively very similar.
To this end, let us consider light propagating in a medium of refractive
index ${\cal N}= 1+ \epsilon$, with $0 \leq \epsilon \ll 1$, and the
effective space-time metric $g^{\mu\nu}=g^{\mu\nu}({\cal N})$ which
should be replaced into the relation $g^{\mu\nu}p_\mu p_\nu=0$.
At the quantum level, this metric was derived by
Jauch and Watson \cite{jauch} when quantizing the electromagnetic
field in a dielectric. They observed that the formalism introduces a
preferred reference system, where the photon energy does
not depend on the direction of light propagation, and which ``is usually taken as the system for which the medium is
at rest''. This conclusion is obvious in Special Relativity, where
there is no preferred system, but less obvious here, where
an isotropic propagation is only assumed when {\it both} medium and observer are at rest
in $\Sigma$.

To be more specific, let us place this medium in two identical optical
resonators, namely resonator 1, which is at rest in $\Sigma$, and
resonator 2, which is at rest in an arbitrary frame $S'$. Let us also introduce
$\pi_\mu\equiv ( {{E_\pi}\over{c}},{\bf \pi }) $, to indicate the
light 4-momentum for $\Sigma$ in his cavity 1, and $p_\mu\equiv (
{{E_p}\over{c}},{\bf p})$, to indicate the analogous 4-momentum of
light for $S'$ in his cavity 2. Finally let us define by
$g^{\mu\nu}$ the space-time metric used by $S'$ in the relation
$g^{\mu\nu}p_\mu p_\nu=0$ and by
\begin{equation}
\label{metricsigma}\gamma^{\mu\nu}={\rm diag}({\cal N}^2,-1,-1,-1)
\end{equation}
the metric which $\Sigma$ adopts in the analogous relation
$\gamma^{\mu\nu}\pi_\mu\pi_\nu=0$ and which produces the isotropic
velocity $c_\gamma=E_\pi/|{\bf \pi}|={{c}\over{{\cal N}}}$.

The peculiar view of Special Relativity is that no
observable difference can exist between two reference systems that are in uniform translational motion. Instead, with a
preferred frame $\Sigma$, as far as light propagation is concerned, this physical equivalence is only assumed in
the ideal ${\cal N}=1$ limit. In fact, for ${\cal N}\neq 1$, where light gets absorbed
and then re-emitted, the fraction of refracted light could keep
track of the particular motion of matter with respect to $\Sigma$
and produce, in a frame $S'$ where matter is at rest, a $\Delta \bar{c}_\theta \neq 0$. Likewise, assuming that
the solid parts of cavity 2 are at rest in the inertial frame $S'$
no longer implies that the medium which stays inside, e.g. a gas, is in thermodynamic equilibrium.
Thus, one should keep an open mind and exploit the implications of the basic
condition \begin{equation} \label{limitingintro} g^{\mu\nu}({\cal N}=1)=
\gamma^{\mu\nu}({\cal N}=1)=\eta^{\mu\nu}\end{equation} where
$\eta^{\mu\nu}$ is the Minkowski tensor. This standard equality
amounts to introduce a transformation matrix, say $A^{\mu}_{\nu}$, which produces $g^{\mu\nu}$
from the reference metric $\gamma^{\mu\nu}$ and such that
\begin{equation}
\label{vacuum}
g^{\mu\nu}({\cal N}=1)= A^{\mu}_{\rho}A^{\nu}_{\sigma}\gamma^{\rho\sigma}  ({\cal N}=1)= A^{\mu}_{\rho}A^{\nu}_{\sigma}\eta^{\rho\sigma}
=\eta^{\mu\nu}
\end{equation}
This relation is strictly valid for ${\cal
N}=1$. However, by continuity, one is driven to conclude that an
analogous relation between $g^{\mu\nu}$ and $\gamma^{\mu\nu}$ should
also hold in the $\epsilon \to 0$ limit. The crucial point is that the chain
in (\ref{vacuum}) does not fix uniquely $A^{\mu}_{\nu}$. In
fact, it is fulfilled either by choosing the identity matrix, i.e.
$A^{\mu}_{\nu}=\delta^{\mu}_{\nu}$, or by choosing a Lorentz
transformation, i.e. $A^{\mu}_{\nu}=\Lambda^{\mu}_{\nu}$. It thus
follows that $A^{\mu}_{\nu}$ is a two-valued function and there are two possible solutions \cite{plus}-\cite{universe}
for the metric in $S'$. In fact, when
$A^{\mu}_{\nu}$ is the identity matrix, we find
\begin{equation}\left[g^{\mu\nu}({\cal N})\right]_1=
\delta^{\mu}_{\rho}\delta^{\nu}_{\sigma}\gamma^{\rho\sigma}=
\gamma^{\mu\nu}\sim \eta^{\mu\nu} + 2\epsilon \delta^\mu_0
\delta^\nu_0\end{equation} while, when $A^{\mu}_{\nu}$ is a Lorentz
transformation, we obtain
\begin{equation} \label{2intro} \left[g^{\mu\nu}({\cal
N})\right]_2= \Lambda^{\mu}_{\rho}
\Lambda^{\nu}_{\sigma}\gamma^{\rho\sigma}\sim \eta^{\mu\nu} +
2\epsilon v^\mu v^\nu
\end{equation} where $v_\mu$ is the
$S'$ 4-velocity, $v_\mu\equiv(v_0,{\bf v}/c)$ with $v_\mu v^\mu=1$.
As a consequence, the equality $\left[g^{\mu\nu}({\cal
N})\right]_1=\left[g^{\mu\nu}({\cal N})\right]_2$ can only hold for
$v^\mu=\delta^\mu_0$, i.e. for ${\bf v}=0$ when $S'\equiv \Sigma$.
Notice that by choosing the first solution $\left[g^{\mu\nu}({\cal
N})\right]_1$, which is implicitly assumed in Special Relativity to
preserve isotropy in all reference frames also for ${\cal N} \neq
1$, we are considering a transformation matrix $A^{\mu}_{\nu}$ which
is discontinuous for any $\epsilon \neq 0$. In fact, all emphasis
on Lorentz transformations depends on enforcing the last equality in
(\ref{vacuum}) for $A^{\mu}_{\nu}=\Lambda^{\mu}_{\nu}$ so that
$\Lambda^{\mu \sigma}\Lambda^{\nu}_{\sigma}=\eta^{\mu\nu}$ and the
Minkowski metric, if valid in one frame,  applies to all equivalent frames.

In conclusion, with a preferred frame $\Sigma$, there may be non-zero $g_{0i}$ in $S'$ which play the role of a velocity field and produce
a small anisotropy of the  two-way velocity \cite{plus}-\cite{universe}
\begin{eqnarray}
\label{twoway00}
       \bar{c}_\gamma(\theta)&=&
       {{ 2  c_\gamma(\theta) c_\gamma(\pi + \theta) }\over{
       c_\gamma(\theta) + c_\gamma(\pi + \theta) }}
       \sim {{c} \over{{\cal N}}}~\left[1-\epsilon\beta^2\left(1 +
       \cos^2\theta\right) \right]\equiv {{c}\over{\bar{\cal N}(\theta)}}
\end{eqnarray}
The only difference with respect to the old ether model
is that  the resulting fractional anisotropy
$\frac { \Delta\bar{c}_\theta}{c} \sim \epsilon (v^2/c^2) $ would be much smaller than the classical prediction
 $\frac { \Delta\bar{c}_\theta}{c}\big|_ {\rm class}\sim  (v^2/2c^2) $. However, this has only a quantitative significance
and has to be decided by experiments.

With this in mind, and addressing to refs.\cite{plus}-\cite{epl} for more details, we will summarize
in Sect.3 the analysis of all data from Michelson-Morley
to the present experiments with optical resonators. As a matter of fact,
once the small residuals are analyzed in a
modern theoretical framework, the $\Sigma-$frame tight to the CMB is
naturally emerging. Sect.4 will finally contain a summary and some
general arguments which may indicate the existence of nearly
instantaneous effects in the physical vacuum.

\section{Vacuum state and its Lorentz invariance}

The discovery of the Higgs boson at LHC has confirmed the basic idea
of Spontaneous Symmetry Breaking (SSB) where particle masses
originate from the particular structure of the vacuum. As
anticipated in the Introduction by 't Hooft's words \cite{thooft},
this means that empty space is actually filled with the elementary
quanta of the Higgs field whose trivial, empty vacuum is
not the lowest-energy state of the theory. In
this section, we will summarize the basic picture of SSB, in the
case of a one-component scalar field $\Phi(x)$
with only a discrete
reflection symmetry $\Phi \to -\Phi$, i.e. no Goldstone bosons. In
spite of its simplicity, this system can already display the general
aspects which are relevant for the problem of a Lorentz-invariant
vacuum state.

Our analysis will be based on the condensation process of {\it
physical} quanta and thus assumes a description of symmetry breaking
as a (weak) first-order phase transition. Namely, SSB occurs when
the renormalized mass squared $m^2_R$ of the quanta of the symmetric
phase is extremely small but still in the {\it physical} region
$m^2_R>0$. In the presence of gauge bosons, this was shown in the
loop expansion by Coleman and Weinberg \cite{CW} long ago. Today,
the same (weak) first-order scenario is now supported by most recent
lattice simulations \cite{lattice1,lattice2,lattice3} of a pure
$\Phi^4$ theory which we adopt as our basic model.

To introduce the issue of Lorentz invariance, let us first recall
that inertial transformations are represented in Hilbert space by
unitary operators which correspond to the Poincar\'e group. This
means a representation of the 10 generators $P_\mu$ and $L_{\mu\nu}$
($\mu$, $\nu$= 0, 1, 2, 3), where $P_\mu$ describe the space-time
translations and $L_{\mu\nu}=-L_{\nu\mu}$ the space rotations and
Lorentz boosts, with commutation relations
\begin{equation} \label{tras1} [P_\mu,P_\nu]=0 \end{equation}
\begin{equation} \label{boost} [L_{\mu\nu}, P_\rho]=
i\eta_{\nu\rho}P_\mu - i\eta_{\mu\rho}P_\nu
\end{equation} \begin{equation} \label{tras2} [L_{\mu\nu},
L_{\rho\sigma}]= -i\eta_{\mu\rho}L_{\nu\sigma}+
i\eta_{\mu\sigma}L_{\nu\rho}
-i\eta_{\nu\sigma}L_{\mu\rho}+i\eta_{\nu\rho}L_{\mu\sigma}
\end{equation} where again $\eta_{\mu\nu}={\rm diag}(1,-1,-1,-1)$ is the Minkowski tensor.
An exact Lorentz-invariant vacuum has to be annihilated by all 10
generators (see e.g. \cite{cpt}).

These premises are well known but one should also be aware that,
with the exception of low-dimensionality
cases, a construction of the Poincar\'e algebra is only known for the
free-field case. For the interacting theory, at present, one can
only implement it perturbatively. This means that one should start
with a definite free-field limit, and therefore with a unique
vacuum, where the simplest prescription of the Wick, normal ordering allows
for a consistent representation of the commutation relations.

Let us thus consider a system of free, spinless quanta with mass $m$
and energy $E(k)=\sqrt{{\bf k}^2 + m^2}$. In terms of annihilation
and creation operators $a({\bf k})$ and $a^{\dagger}({\bf k})$ of an
empty vacuum $|{\rm o}\rangle$, with $a({\bf k})|{\rm
o}\rangle=\langle{\rm o}|a^{\dagger}({\bf k})=0$, and commutation
relations $[ a({\bf k}), a^{\dagger}({{\bf k}'}) ] = \delta_{{\bf k}
{\bf k'}}$, the required representation of the generators is then
\BE \label{freehamiltonian} P_0 \equiv~ H_2(m)~ = ~:\int \! d^3 x
\left[ \frac{1}{2} \left( \Pi^2 + (\del \Phi)^2 + m^2 \Phi^2 \right)
\right]: ~=\sum_{\bf k} E(k) a^{\dagger}({\bf k)} a({\bf k}) \EE \BE
\label{momentum} P_i = ~~ :\int \! d^3 x
 \Pi(x) \partial_i \Phi(x) :
 ~~=\sum_{\bf k} k_i  a^{\dagger}({\bf k})  a({\bf k}) \EE
\begin{eqnarray}
\label{angularmomentum} L_{ij} = :\int \! d^3 x
[x_i\Pi(x) \partial_j\Phi(x) - x_j\Pi(x) \partial_i\Phi(x)] : =\nonumber\\
 =  \frac{i}{2} \sum_{\bf k} [k_i  a^{\dagger}({\bf k)}  \frac{\partial}{\partial k_j} a({\bf k})
- k_j  a^{\dagger}({\bf k)}
\frac{\partial}{\partial k_i} a({\bf k}) ]
\end{eqnarray}
\begin{eqnarray}
\label{boosti} L_{0i} =  x_0 P_i - \frac{1}{2}:\int \! d^3 x  ~x_i [\Pi^2 + (\del \Phi)^2 + m^2 \Phi^2 ]:=\nonumber\\
=\frac{i}{2} \sum_{\bf k}  E(k)   a^{\dagger}({\bf k)}   \frac{\partial}{\partial k_i} a({\bf k})
\end{eqnarray}
With the above expressions, the Poincar\'e algebra is reproduced and
the empty vacuum $|{\rm o}\rangle$ is annihilated by all 10
generators. As such, the general requirements for an exact Lorentz
invariant vacuum are fulfilled. In particular, notice the crucial
role of the zero-energy condition $ P_0|{\rm o}\rangle=0$ which,
from the commutation relations $[P_i, {L}_{0i}]= iP_0 $ (no
summation over $i$), is needed for consistency with $ P_i|{\rm
o}\rangle=0$ and ${L}_{0i}|{\rm o}\rangle=0$.

Let us now introduce the interaction and consider the limit of a
very weakly coupled $g\Phi^4$ theory, i.e. with a coupling $g$ in
the range $ 0<g \ll 1$. In this case, which is the typical example
of non-linear QFT with polynomial interaction $P(\Phi)= g \Phi^4$,
there are two basically different options. In a first approach, one
could try to introduce a suitable de-singularized operator, say
$::P(\Phi(x))::$, which extends the standard normal ordering of the
free-field case so that
$\langle\Psi_0|::P(\Phi(x))::|\Psi_0\rangle=0$ in the true vacuum
state $|\Psi_0\rangle$. This type of approach, which has been
followed by very few authors, was discussed by Segal \cite{segal}.
His conclusion was that $::P(\Phi(x))::$ is not well-defined until
the physical vacuum is known, but, at the same time, the physical
vacuum also depends on the definition given for $::P(\Phi(x))::$.
From this type of circularity Segal was deducing that, in general,
in such a nonlinear QFT, the physical vacuum will {\it not} be
invariant under the full Lorentz symmetry of the underlying
Lagrangian density.

A second approach, followed nowadays by most authors, is instead to
consider $g\Phi^4$ theory in the framework of a perturbative
Renormalization-Group approach. In this case, $g \equiv g(\mu)$ should be
understood as the running coupling constant at a variable mass scale
$\mu$ and, in its value, would carry the information on the
asymptotic pair $(g_0, \Lambda_s)$ where $g_0$ is the bare coupling
at some minimum locality scale fixed by the ultraviolet cutoff
$\Lambda_s$. Clearly, with a finite $\Lambda_s$, one is explicitly
breaking Lorentz symmetry. However, by the generally accepted
`triviality' of $g\Phi^4$ theory in 4 space-time dimensions, one
finds that $g(\mu)\sim \ln^{-1}(\Lambda_s/\mu) \to 0^+$ for $\Lambda_s \to \infty$, whatever the value
of $g_0$. By defining $m$ the mass of the scalar quanta in the symmetric phase
$\langle \Phi \rangle=0$, one can then assume that, for $\mu \sim m$,  $g$ is so small
(or equivalently the cutoff $\Lambda_s$ is so large) that one can
meaningfully obtain $|\Psi_0\rangle$ by perturbing
around the previous free-field vacuum $ |{\rm o}\rangle$. Likewise, with a formally
Lorentz-invariant interaction such as $\Phi^4$, it should be possible to
construct a representation of the Poincar\'e algebra
Eqs.(\ref{tras1}),(\ref{boost}),(\ref{tras2}) which holds {\it to
any finite order in $g$} and goes smoothly into the previous
free-field structure for $\Lambda_s \to \infty$. Finally, as in the free-field
case, since the true vacuum $|\Psi_0\rangle$ is always assumed to
have zero spatial momentum, from the commutator $[P_i, {L}_{0i}]=
iP_0 $, its invariance under boosts will also require to implement the
zero-energy condition in the perturbative expansion.

This general strategy is
briefly sketched below with a Hamiltonian
 \BE\label{fullhamiltonian2} P_0\equiv
H =H_2(m) + H_I + \Delta E\EE
which beside the interaction
\BE\label{HI} H_I= : \int \!
d^3 x~\frac{g}{4!} \Phi^4: \EE
should also include an additive constant
$\Delta E$  given as a power series in $g$ \BE \label{DeltaE} \Delta E= g \Delta E(1) +
g^2 \Delta E(2) +...\EE
with coefficients $\Delta E(1)$, $\Delta
E(2)$... determined by imposing that the true ground state \BE
\label{true} |\Psi_0 \rangle = |{\rm o}\rangle+ g |{\rm o}(1)\rangle
+ g^2|{\rm o}(2)\rangle +...\EE has exactly zero energy to any
finite order in $g$ \BE \label{zero} P_0|\Psi_0 \rangle=0\EE
Again $|{\rm o}\rangle$ is the vacuum of the free-field $H_2(m)$,
with $H_2(m)|{\rm o}\rangle=0$, so that if, with a short-hand
notation, we denote by $|{\rm n}\rangle$ its higher eigenstates,
i.e.  $H_2(m)|{\rm n}\rangle=E_n|{\rm n}\rangle$ with
$E_n>0$, we find $\Delta E(1)=0$ and the first few relations \BE
g|{\rm o}(1)\rangle = -\sum_{n\neq 0}~ \frac{|{\rm
n}\rangle\langle{\rm n}|H_I|{\rm o}\rangle }{E_n} \EE \BE g^2|{\rm
o}(2)\rangle = \sum_{m\neq 0}\sum_{n\neq 0}~ \frac{|{\rm n}\rangle
\langle{\rm n}|H_I|{\rm m}\rangle\langle{\rm m}|H_I|{\rm
o}\rangle }{E_nE_m} \EE \BE g^2\Delta
E(2)=\sum_{n\neq 0}~ \frac{\langle{\rm o}|H_I|{\rm
n}\rangle\langle{\rm n}|H_I|{\rm o}\rangle}{E_n}\EE The issue of
Lorentz invariance would finally require the analysis of the boost
generators $L_{0i}$ which should also be re-defined in perturbation
theory \cite{Stefanovich,Glazek} by starting from the free-field form $L_{0i}(g=0)$ in
Eq.(\ref{boosti}), say \BE \label{boostig}
L_{0i}(g)= L_{0i}(g=0) + g \Delta L_{0i}(1) + g^2 \Delta L_{0i}(2) +
...\EE In this way, with a zero-energy vacuum
state $|\Psi_0 \rangle$, and assuming that the Poincar\'e algebra
Eqs.(\ref{tras1}),(\ref{boost}),(\ref{tras2}) and the condition
$L_{0i}(g) |\Psi_0 \rangle=0$ can be fulfilled to any finite
order $g^n$, i.e. up to $g^{n+1}$ terms, the Lorentz invariance of
$|\Psi_0\rangle$ can be considered exact.

\begin{figure}[h]
\begin{center}
\includegraphics[width=10.5 cm]{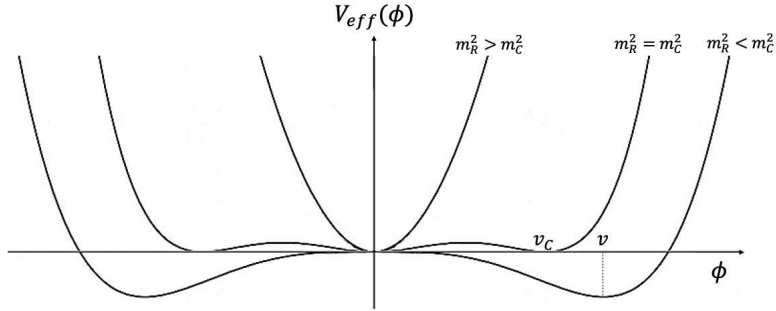}
\end{center}
\caption{\it A schematic profile of the effective potential where SSB is a 1st-order phase transition.} \label{1st-order}
\end{figure}

Let us now consider the phenomenon of SSB. Here, for constant field
configurations, the basic quantity is the effective potential
$V_{\rm eff} (\phi)$. Within the class of normalized quantum states
$\langle\Psi|\Psi\rangle=1$, this is defined in the infinite volume
limit ${\vol}\to \infty$ as \BE \label{Veff} V_{\rm eff} (\phi)=
\frac{1}{\vol}\cdot {\rm min}_\Psi \langle\Psi |H| \Psi\rangle \EE
with the condition \BE \label{phi1} \langle\Psi |\Phi| \Psi\rangle =
\phi \EE By expanding around $\phi=0$, the quadratic shape of the
effective potential gives the renormalized mass of the
symmetric-phase quanta \BE \label{m2R} m^2_R=V''_{\rm eff}
(\phi)\Big|_{\phi=0}\EE given as a power serious expansion
 \BE \label{m2R2} m^2_R= m^2 + g^2\Delta m^2(2) + g^3 \Delta m^2(3) +...\EE
In turn, from the connection between
variational principle and the eigenvalue equation for the
Hamiltonian, for a translational invariant field configuration, the
previous zero-energy condition of the symmetric ground state at
$\langle\Psi_0 |\Phi| \Psi_0\rangle=0$ implies \BE
\label{zeroeff}V_{\rm eff} (\phi=0)=0\EE  to any finite order in
$g$.

The conventional second-order picture of SSB assumes that no
non-trivial minima can exist if $m^2_R \ge 0$. Instead, as indicated
by most recent lattice simulations \cite{lattice1,lattice2,lattice3}
of a pure $\Phi^4$ theory, we will adopt the view of SSB as a (weak)
first-order phase transition. This means that, with the same
Hamiltonian Eq.(\ref{fullhamiltonian2}), before reaching the
massless $m^2_R= 0$ limit, for some very small $m^2_R= m^2_c$ there will arise
a first pair of vacua, say with $\langle \Phi \rangle= \pm v_c$,
which have the same zero energy as the empty vacuum at $\langle \Phi
\rangle=0$ see Fig. \ref{1st-order}. Therefore the critical mass is defined as that value \BE
\label{m2c} m^2_c=V''_{\rm eff} (\phi)\Big|_{\phi=0}\EE for which
there are $three$ vacuum states with the same energy \BE
\label{equal} 0=V_{\rm eff} (\phi=0)=V_{\rm eff} (\phi=\pm v_c)\EE
On the other hand, for $0\leq m^2_R< m^2_c$, SSB takes place and the
energy of the two degenerate vacua, say $|\Psi_{\pm}\rangle$ with
$\langle\Psi _{\pm}| \Phi |\Psi_{\pm}\rangle= \pm v$, will
definitely be lower than its value at $\langle \Phi \rangle=0$
\BE\label{unequal} V_{\rm eff} (\phi=\pm v) < V_{\rm eff}
(\phi=0)=0\EE Explicit calculations, either in the loop expansion
\cite{mech} or variational \cite{ciancitto}, indicate that $V_{\rm
eff} (\phi=\pm v)$ has a non analytic $\exp(-1/g)$ behaviour. This
reflects the basic non perturbative nature of SSB which cannot be
found to any finite order in $g$.

Here we emphasize a few aspects:

~~i) to understand intuitively the first-order nature of the phase
transition, the crucial observation is that the quanta of the
symmetric phase, besides the +$g \delta^{(3)}(\bf r)$
repulsion, also feel a $-g^2 \frac {e^{-2 m r}}{r^3}$
attraction \cite{mech} which shows up at the one-loop level and
whose range becomes longer and longer in the $m \to 0$ limit. A
calculation of the energy density in the dilute-gas approximation,
which is equivalent to the one-loop effective potential, indicates
that for very small $m$ the attractive tail dominates.
Higher-order corrections simply renormalize the strength of these
two basic effects \cite{mech} whose interplay explains the
instability of the symmetric phase producing a
$\langle\Phi\rangle\neq 0$

~~ii) the field fluctuations around each of the two non-symmetric
vacua are conveniently described by the shifted field $h(x)\equiv
\Phi(x)-\langle \Phi \rangle$, with $\langle h\rangle=0$ by
definition. In the realistic case of the Standard Model with a
SU(2)xU(1) symmetry, the lowest $h-$field excitation has mass
$m_h\sim$ 125 GeV as observed at the Large Hadron Collider of CERN.
Notice that this $m_h$ is conceptually different from the $m_R$
discussed above, the two mass parameters describing the quadratic
shape of the effective potential at two different values of $\phi$,
namely \BE m^2_h=V''_{\rm eff} (\phi)\Big|_{\phi=\pm v}\EE

~~iii) strictly speaking, Wightman's axioms \cite{cpt} require a
unique vacuum state. The usual way out is that the two states
$|\Psi_{\pm}\rangle$, giving the absolute minima of the effective
potential, have zero overlap $\langle\Psi _{-}|\Psi_{+}\rangle\to 0
$ in the infinite-volume ${\vol}\to \infty$ \cite{weinbergQFT}

~~iv) due to the equivalence between variational method and
eigenvalue equation, the two spontaneously broken vacua
$|\Psi_{\pm}\rangle$, at the minima of the effective potential, are
the lowest eigenstates of the Hamiltonian. However, by assuming
Eq.(\ref{zero}), their energy {\it cannot} be zero. Then, from
$[P_i, {L}_{0i}]= iP_0 $, $ P_i |\Psi_{\pm}\rangle=0$ and
$P_0|\Psi_{\pm}\rangle \neq 0$ it follows ${L}_{0i}
|\Psi_{\pm}\rangle \neq 0$ so that the two states
$|\Psi_{\pm}\rangle$ cannot be Lorentz invariant
\cite{epjc,dedicated,foop}.

Of course, concerning iv), we could define a different free-field
limit to preserve the Lorentz invariance of the two
$|\Psi_{\pm}\rangle$. To this end, one should first express the
field as $\Phi(x)=h(x)+ \langle \Phi \rangle$ so that the original
$\Phi^4$ term will also produce a cubic term $g\langle \Phi \rangle
h^3$ which reflects the interaction of the $h$-field fluctuation
with the vacuum condensate $\langle \Phi \rangle$. Then, $ h(x)$
should be quantized in terms of new annihilation and creation
operators, say $b({\bf k})$ and $b^{\dagger}({\bf k})$ for
$|\Psi_{+}\rangle$ and $c({\bf k})$ and $c^{\dagger}({\bf k})$ for
$|\Psi_{-}\rangle$, so that $b({\bf k})|\Psi_{+}\rangle=0$ and
$c({\bf k})|\Psi_{-}\rangle=0$. This means that one should select
one of the two non-symmetric vacuum states, for instance
$|\Psi_{+}\rangle$, and define the normal ordering procedure, in
terms of $b({\bf k})$ and $b^{\dagger}({\bf k})$. After having
re-defined the free-field limit, the perturbative expansion should
also be re-formulated because, now, there is a new dimensionful
coupling constant $g\langle \Phi \rangle$ \footnote{The presence of
the cubic $g\langle \Phi \rangle h^3$  interaction should not be
overlooked. In fact, in the infrared region, it induces a strong
coupling between bare $b^{\dagger}|\Psi_{+}\rangle$ and
$b^{\dagger}b^{\dagger}|\Psi_{+}\rangle$ components in the Fock
space of the broken-symmetry phase \cite{mpla11}. The net result is
that, in the ${\bf k}\to 0$ limit, the effective 1-(quasi)particle
spectrum deviates sizeably from the spectrum of the bare
$b^{\dagger}|\Psi_{+}\rangle$ states. }. Finally, the additive
constant $\Delta E$ should now be determined by requiring zero
energy for $|\Psi_{+}\rangle$ with a change in the Hamiltonian
Eq.(\ref{fullhamiltonian2}). Correspondingly, the previous symmetric
vacuum $|\Psi_0 \rangle$, would now be higher of the same amount
associated with $V_{\rm eff} (\phi=\pm v)$.

With these two different procedures, one has to choose between the
Lorentz invariance of the two degenerate minima $|\Psi_{\pm}\rangle$
and the Lorentz invariance of the original $|\Psi_0 \rangle$. In
particular, with the alternative re-arrangement described above, we
are implicitly admitting that, even in the simplest case of a
one-component, massive $\Phi^4$ theory with  $g \to 0^+$, there is
no way to start from the free-field vacuum $|{\rm o}\rangle$ and
preserve, in perturbation theory, the basic Lorentz symmetry
embodied in the operatorial structure
Eqs.(\ref{freehamiltonian})-(\ref{boosti}) \footnote{One may object
that this conflict is just a consequence of describing SSB  as a
(weak) first-order phase transition. Apparently, in the standard
second-order picture, where no meaningful quantization in the
symmetric phase is possible, there would be no such a problem.
However this is illusory. In fact,  as previously recalled, the same
(weak) first-order scenario is found, within the conventional loop
expansion \cite{CW}, when studying SSB in the more realistic case of
complex scalar fields interacting with gauge bosons. This
first-order scenario is at the base of
 't Hooft's description \cite{thooft} of the physical vacuum as a Bose condensate of real, physical Higgs quanta (not of tachions with imaginary mass).
Thus the problem goes beyond the simplest $\Phi^4$ model considered here and reflects the general constraint imposed by SSB on the
renormalized mass parameter of the scalar fields in the symmetric phase. }.
Perhaps, a way out could be to impose the
condition $m^2_R=m^2_c$ which fixes the mass of the
symmetric phase at the particular critical value where
Eq.(\ref{equal}) holds true and the three local minima have all the
same zero energy (a choice which, however, would imply the
meta-stability of the broken-symmetry phase).

By discarding the particular case $m^2_R=m^2_c$, we will then return
to our original choice of a Lorentz invariant $|\Psi_0\rangle$. The point is
that the resulting Lorentz-non-invariance of the two non-symmetric minima
$|\Psi_{\pm}\rangle$ has an intuitive physical meaning. Indeed, it
reflects the macroscopic occupation of the same quantum state
by the basic elementary quanta of $\Phi^4$, those described by
$a({\bf k})$ and $a^{\dagger}({\bf k})$, and which behave as hard
spheres at the asymptotic cutoff scale $\Lambda_s$.

To this end, let us consider the field expansion in terms of $a({\bf
k})$ and $a^{\dagger}({\bf k})$\BE \Phi({\bf x})= \sum_{\bf k}
\frac{1} {\sqrt{2 \vol E(k)} } \left[ a({\bf k}) e^{i {\bf k\cdot
x}} + a^{\dagger}({\bf k})e^{-i {\bf k\cdot x}}\right] \EE and
introduce the number of field quanta through the operator \BE
\label{nop} \hat{\num} = \sum_{\bf k} a^{\dagger}_{\bf k} a_{\bf k}
\EE For a temperature $T= 0$, following 't Hooft \cite{thooft}, see
also \cite{mech}, SSB corresponds to a macroscopic number $\num$ of
quanta in the state ${\bf k}=0$ of some reference frame $\Sigma$. In
this case, where $\hat{\num} \sim a^{\dagger}_0 a_0$, one can
effectively consider $a_0$ as a c-number $\sqrt{\num}$ with \BE \phi
= \langle \Phi \rangle \sim \frac{1}{\sqrt{2 \vol m}}~ 2 a_0 \sim
\sqrt{\frac{2\num}{\vol m}} \EE  a density $n \equiv \num/\vol$
(with $\num \to \infty$,  $\vol \to \infty$ at fixed $n$) given by
\BE \label{neq} n \sim \half m \phi^2\EE and mass density $\rho_m=
(m \num/\vol)$ reproducing the quadratic term $ \half m^2 \phi^2$ in
the potential.

In this formalism, it will be natural to adopt a particular notation
for each of the two degenerate vacua $|\Psi_{\pm}\rangle$, say
$|\Psi^{(\Sigma)}_{\pm}\rangle$, which defines the vacuum assignment
for that observer which is at rest in $\Sigma$. As we have seen, if
these states have a non-zero energy $E_0 <0$, and thus are not
Lorentz invariant, boost operators ${U}'$, ${U}''$,..will transform
non-trivially $|\Psi^{(\Sigma)}_{\pm}\rangle$ into new states $|
\Psi'_{\pm}\rangle$, $| \Psi''_{\pm}\rangle$,... appropriate to
moving observers $S'$, $S''$,..For instance, by defining the boost
operator $U'=e^{i\lambda'{L}_{01}}$ one finds
\begin{equation}\label{trasf1} | \Psi'_\pm \rangle=
e^{i\lambda'{L}_{01}}|\Psi^{(\Sigma)}_\pm \rangle \end{equation} so
that, by using the relations \begin{equation} \label{po1}
e^{-i\lambda'{L}_{01}}~{P}_1~
e^{i\lambda'{L}_{01}}=\cosh\lambda'~{P}_1 + \sinh\lambda'~{P}_0
\end{equation} \begin{equation} \label{po2} e^{-i\lambda'{L}_{01}}~{P}_0
~e^{i\lambda'{L}_{01}}=\sinh\lambda'~{P}_1 + \cosh\lambda'~{P}_0
\end{equation} we obtain \begin{equation} \label{po3}
 {{P}_1} |\Psi'_{\pm}\rangle=E_0\sinh\lambda' |\Psi'_{\pm}\rangle~~~~~~~~~~~~~~~
 {{P}_0} |\Psi'_{\pm}\rangle=E_0\cosh\lambda'|\Psi'_{\pm}\rangle \end{equation}
Thus, the physical, realized form of relativity would now contain a
preferred frame $\Sigma$ with zero spatial momentum. In a generic
moving system $S'$, one would instead find  \footnote{This contrasts
with the approach based on an {\it energy-momentum tensor} of the
form \cite{zeldovich,weinbergreview}
\begin{equation}\label{zeld}  \langle \Psi^{(\Sigma)}_{\pm}| {W}_{\mu\nu}  |\Psi^{(\Sigma)}_{\pm}\rangle
=\rho_v ~\eta_{\mu\nu}\end{equation} $\rho_v$ being a space-time
independent constant. In fact, from $ \langle \Psi'_{\pm}|
{W}_{\mu\nu}  |\Psi'_{\pm}\rangle
=\Lambda^{\sigma}{_\mu}\Lambda^{\rho}{_\nu} ~ \langle
\Psi^{(\Sigma)}_{\pm}| {W}_{\sigma\rho}
|\Psi^{(\Sigma)}_{\pm}\rangle$ and  Eq.(\ref{zeld}) one finds
$\langle\Psi'_{\pm}| {W}_{0 i}  |\Psi'_{\pm}\rangle=0$ so that
$\langle \Psi'_{\pm}|{{P}_i} |\Psi'_{\pm}\rangle = 0 $. However,
Eq.(\ref{zeld}) does not correspond to the idea of
$|\Psi^{(\Sigma)}_\pm \rangle$ as the lowest-energy eigenstate
which, as anticipated, is implicit in the notion of a minimum of
$V_{\rm eff} (\phi)$. This is important if we are interested in the
Lorentz invariance of the vacuum. Within the Poincar\'e algebra,
this is a well defined problem requiring $|\Psi^{(\Sigma)}_\pm
\rangle$ to be annihilated by the boost generators \BE
{L}_{0i}=-\int d^3x~(x_i{W}_{00}-x_0 {W}_{0i}) \EE If
$|\Psi^{(\Sigma)}_\pm \rangle$ is an eigenstate of the Hamiltonian,
then an eigenvalue $E_0= 0$ is needed to obtain ${L}_{0i}
|\Psi^{(\Sigma)}_\pm \rangle = 0$. Instead, Eq.(\ref{zeld}) amounts
to $\langle  \Psi^{(\Sigma)}_{\pm}| {L}_{0i}
|\Psi^{(\Sigma)}_{\pm}\rangle =0$. Thus, no surprise that one can
run into contradictory statements.} \BE \label{nonzeroP_i}
{{P}_i}|\Psi'_{\pm}\rangle=E_0 b_i|\Psi'_{\pm}\rangle ~~~~~~~~~
{{P}_0}|\Psi'_{\pm}\rangle=E_0 a |\Psi'_{\pm}\rangle ~~~~~~~
a^2-b_ib_i=1 \EE We observe that, traditionally, with the exception
of Chiao's \cite{Chiao} mentioned `conceptual tensions', SSB was
never believed to be in a potential conflict with Einstein
relativity. The motivation being, perhaps, that the mean properties
of the condensed phase are summarized into the vacuum expectation
value $\langle\Phi\rangle$ of the Higgs field which transforms as a
world scalar under the Lorentz group. However, this does not imply
that the vacuum state itself is {\it Lorentz invariant}. Lorentz
transformation operators ${U}'$, ${U}''$,...could transform non
trivially the reference vacuum states
$|\Psi^{(\Sigma)}_{\pm}\rangle$ and, yet, for any Lorentz scalar
operator ${S}$, i.e. for which $S=(U')^\dagger S U'=(U'')^\dagger S
U''...$, one would find \BE \langle\Psi^{(\Sigma)}_{\pm}| {S}
|\Psi^{(\Sigma)}_{\pm}\rangle =\langle\Psi'_{\pm}|
{S}|\Psi'_{\pm}\rangle=\langle\Psi''_{\pm}|
{S}|\Psi''_{\pm}\rangle...\end{equation} Another aspect, which is
always implicitly assumed but very seldom spelled out, concerns the
condition $V_{\rm eff}(\phi=0)=0$. This is usually interpreted as a
matter of convention, as if one were actually computing $\Delta
V_{\rm eff} (\phi)\equiv V_{\rm eff} (\phi)- V_{\rm eff} (\phi=0)$.
However, starting from this apparently innocent assumption, many
authors have raised the problem of the non-zero cosmological
constant in Einstein's field equations which is generated by SSB.
This problem has only a definite meaning if the condition $V_{\rm
eff} (\phi=0)=0$ is not arbitrary but is actually assumed from the
start for consistency. With our previous analysis of the symmetric
vacuum, we can now understand the motivations of this implicit
assumption. Starting from the free-field structure in
Eqs.(\ref{freehamiltonian})- (\ref{boosti}), the condition
Eq.(\ref{zeroeff}) expresses the requirement of having a zero-energy
vacuum at $\phi=0$ and, therefore, of preserving its Lorentz
invariance in the interacting theory. Still, near the critical mass,
where Eq.(\ref{equal}) holds true, the induced cosmological constant
could be made arbitrarily small \footnote{By extending the
Poincar\'e algebra, a remarkable case which fulfills the zero-energy
condition exactly, is that of an unbroken supersymmetric theory.
This is because the Hamiltonian $H \sim \bar{Q}^\alpha Q^\alpha$ is
bilinear in the supersymmetry generators $Q^\alpha$. Therefore an
exact supersymmetric state, for which $Q^\alpha|\Psi\rangle=0$, has
automatically zero energy. At present, however, an unbroken
supersymmetry is not phenomenologically acceptable.}.

Truly enough, the previous arguments are rather formal and give no
information on the preferred frame $\Sigma$ tight to the reference
vacua $|\Psi^{(\Sigma)}_{\pm}\rangle$. For this reason, in the
following Sect.3, we will turn our attention to experiments and try
to understand if $\Sigma$ really exists and if, eventually, is tight
to the CMB as assumed in refs.
\cite{scarani,cocciaro,salart,bancal,cocciaro2}.

\section{The basics of the ether-drift experiments}

\subsection{Which preferred frame?}



Looking for the preferred $\Sigma-$frame, the natural candidate is
the reference system where the temperature of the CMB looks exactly
isotropic or, more precisely, where the CMB  Kinematic Dipole
\cite{yoon} vanishes. This dipole is in fact a consequence of the
Doppler effect associated with the motion of the Earth ($\beta=V/c)$
\BE T(\theta)={{T_o\sqrt{1-\beta^2}}\over{1- \beta \cos \theta} }
\EE Accurate observations with satellites in space \cite{smoot} have
shown that the measured temperature variations correspond to a
motion of the solar system described by an average velocity $V\sim
370$ km/s, a right ascension $\alpha \sim 168^o$ and a declination
$\gamma\sim -7^o$, pointing approximately in the direction of the
constellation Leo. This means that, if one sets $T_o \sim $ 2.725 K
and $\beta\sim 0.00123$, there are angular variations of a few
millikelvin \BE \label{CBR}\Delta T^{\rm CMB}(\theta) \sim T_o \beta
\cos\theta \sim \pm 3.36 ~{\rm mK} \EE which represent by far the
largest contribution to the CMB anisotropy.

Therefore, one may ask, could the reference system with vanishing
CMB dipole represent a fundamental preferred frame for relativity as
in the original Lorentzian formulation? The standard answer is that
one should not confuse these two concepts. The CMB is a definite
medium and, as such, sets a rest frame where the dipole anisotropy
is zero. Our motion with respect to
this system has been detected but there is no
contradiction with Special Relativity.

Though, to good approximation, this kinematic dipole arises by combining
the various forms of peculiar motion which are
involved (rotation of the solar system around the center of the
Milky Way, motion of the Milky Way toward the center of the Local
Group, motion of the Local Group of galaxies in the direction of the
Great Attractor...) \cite{smoot}. Thus, if one could switch-off the
local inhomogeneities which produce these peculiar forms of motion,
it is natural to imagine a global frame of rest associated with the
Universe as a whole. A vanishing CMB dipole could then just {\it
indicate} the existence of this fundamental system $\Sigma$ that we
may conventionally decide to call `ether' but the cosmic radiation
itself would not {\it coincide} with this form of ether. Due to the
group properties of Lorentz transformations, two observers S' and
S'', moving individually with respect to $\Sigma$, would still be
connected by a Lorentz transformation with relative velocity
parameter fixed by the standard relativistic composition rule
\footnote{We ignore here the subtleties related to the Thomas-Wigner
spatial rotation which is introduced when considering two Lorentz
transformations along different directions, see e.g. \cite{ungar,
costella, kanevisser}.}. But, as anticipated in the Introduction,  ultimate consequences would be far
reaching.

The answer cannot be found on pure theoretical grounds and this is
why one looks for a small anisotropy of the two-way velocity of
light $\frac { \Delta\bar{c}_\theta}{c} \neq 0$ in the Earth
laboratory. Here, the general consensus is that no genuine ether
drift has ever been observed, all measurements (from
Michelson-Morley to the most recent experiments with optical
resonators) being seen as a long sequence of null results, i.e.
typical instrumental effects in experiments with better and better
systematics (see e.g. Figure 1 of ref.\cite{nagelnature}).

However, this is not necessarily true. In the original measurements,
light was propagating in gaseous systems (air or helium at
atmospheric pressure) while now, in modern experiments, light
propagates in a high vacuum or inside solid dielectrics. Therefore,
in principle, the difference with the modern experiments might not
depend on the technological progress only but also on the different
media that are tested thus preventing a straightforward comparison.
This is even more true if one takes into account that, in the past,
greatest experts (as Hicks and Miller) have seriously questioned the
traditional null interpretation of the very early measurements. The
observed `fringe shifts', although much smaller than the predictions
of classical physics, were often non negligible as compared to the
extraordinary sensitivity of the interferometers. It is then conceivable that, in some
alternative framework, the small residuals could acquire a
physical meaning. As a definite example, in the following Subsect.
3.2, we will summarize the theoretical scheme of
refs.\cite{plus,plus2,book,universe} starting with the old experiments.
The modern experiments will be considered in Subsect.3.3.

\subsection{The old experiments in gaseous media}

In the old experiments in gases (Michelson-Morley, Miller,
Tomaschek, Kennedy, Illingworth, Piccard-Stahel,
Michelson-Pease-Pearson, Joos) \cite{mm}-\cite{joos}, with
refractive index ${\cal N}=1 + \epsilon$, the velocity of light in
the interferometers, say $c_\gamma$, was not the same parameter $c$
of Lorentz transformations. Hence, assuming their exact validity,
deviations from isotropy could only be due to the small fraction
of refracted light which keeps track of the velocity of matter
with respect to $\Sigma$ and produces a direction-dependent
refractive index. As anticipated in the Introduction, from symmetry arguments valid in the $\epsilon \to
0$ limit \cite{plus,plus2,book,universe}, one would then expect a
two-way velocity
\begin{eqnarray}
\label{twoway0}
       \bar{c}_\gamma(\theta)&=&
       {{ 2  c_\gamma(\theta) c_\gamma(\pi + \theta) }\over{
       c_\gamma(\theta) + c_\gamma(\pi + \theta) }}
       \sim {{c} \over{{\cal N}}}~\left[1-\epsilon\beta^2\left(1 +
       \cos^2\theta\right) \right]\equiv {{c}\over{\bar{\cal N}(\theta)}}
\end{eqnarray}
with an effective $\theta-$dependent refractive index \footnote{A conceptual detail concerns the relation of the gas refractive
index ${\cal N}$, as introduced in Eq.(\ref{metricsigma}), to the
experimental quantity ${\cal N}_{\rm exp}$ which is extracted from
measurements of the two-way velocity in the Earth laboratory. By
assuming a $\theta-$dependent refractive index as in
Eq. (\ref{nbartheta})
one should thus define ${\cal N}_{\rm exp}$
by an angular average, i.e.
$ {{c}\over{ {\cal N}_{\rm exp} }}\equiv
\langle {{c}\over{\bar{\cal N}(\theta)}} \rangle_\theta= {{c}\over{
{\cal N}  }} ~\left[1-{{3}\over{2}} ({\cal N} -1)\beta^2\right]$.
One can then determine the unknown value
${\cal N}  \equiv {\cal N}(\Sigma)$ (as if the container of the gas
were at rest in $\Sigma$), in terms of the experimentally known
quantity ${\cal N}_{\rm exp}\equiv{\cal N}({\rm Earth})$ and of $v$. As discussed in
refs. \cite{plus}-\cite{universe}, for $v\sim 370$ km/s, the difference of the two quantities is well below the
experimental accuracy and, for all practical purposes,can be neglected.} \BE
\label{nbartheta}\bar{\cal N}(\theta)\sim {\cal N}
~[1+\epsilon\beta^2 (1 + \cos^2\theta)] \EE and a fractional
anisotropy
\begin{equation} \label{bbasic2new} {{\Delta \bar{c}_\theta }
\over{c}}
={{\bar{c}_\gamma(\pi/2+\theta)-\bar{c}_\gamma(\theta)}\over{c}}\sim
     \epsilon~\beta^2
       \cos2\theta \end{equation}
In the above relations, $\beta\equiv v/c$, with $v$ and $\theta=0$
indicating respectively the magnitude and the direction of the drift
in the plane of the interferometer.
\begin{figure}[h]
\begin{center}
\includegraphics[width=8 cm]{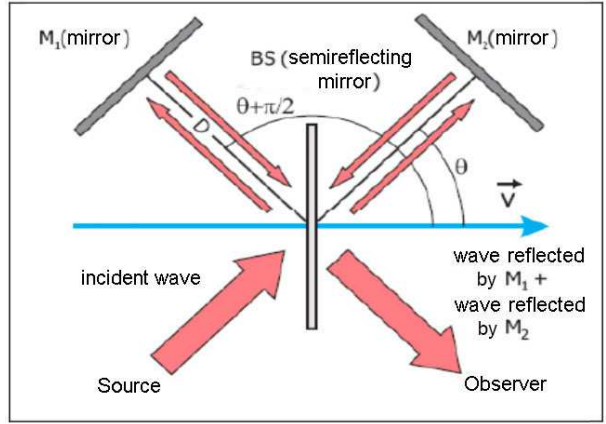}
\end{center}
\caption{\it A schematic illustration of the Michelson
interferometer. Note that, by computing the transit times and the resulting fringe shifts Eq.(\ref{newintro1}), we are assuming the validity
of Lorentz transformations so that the length of a rod does not
depend on its orientation, in the frame $S'$ where it is at rest.} \label{Michinterferometer}
\end{figure}
By introducing the optical path $D$, see
Fig.\ref{Michinterferometer}, and the light wavelength $\lambda$,
this would produce a fringe pattern
\begin{equation} \label{newintro1} {{\Delta
\lambda(\theta)}\over{\lambda}}= {{2D}\over{\lambda}} ~{{\Delta
\bar{c}_\theta } \over{c}}\sim {{2D}\over{\lambda}}~
\epsilon~{{v^2}\over{c^2}}\cos 2\theta
\end{equation}
so that the dragging of light in the Earth frame is described as a
pure 2nd-harmonic effect which is periodic in the range $[0,\pi]$,
as in the classical theory (see e.g. \cite{kennedy}), with the
exception of its amplitude
\begin{equation} \label{a2new}
 A_2={{2 D}\over{\lambda}} ~\epsilon ~{{{v}^2}\over{c^2}}
\end{equation}
This is suppressed by the factor $2\epsilon$ relatively to the
classical amplitude for the orbital velocity  of 30 km/s  \BE \label{a2class}
A^{\rm class}_2={{ D}\over{\lambda}} (\frac{30~{\rm km/s}}{c})^2\EE
This difference could then be re-absorbed into an {\it observable}
velocity which depends on the gas refractive index
\begin{equation}  \label{vobs} v^2_{\rm obs} \sim 2\epsilon v^2 \end{equation}
and is the very small velocity $5-10$ km/s traditionally extracted from
the classical analysis of the early experiments through the relation
\begin{equation}  \label{vobs2} v_{\rm obs} \sim 30 ~{\rm km/s} ~\sqrt{\frac {A^{\rm EXP}_2}{A^{\rm class}_2}}\end{equation}
see e.g. Fig.\ref{miller}.
\begin{figure}[h]
\begin{center}
\includegraphics[width=10 cm]{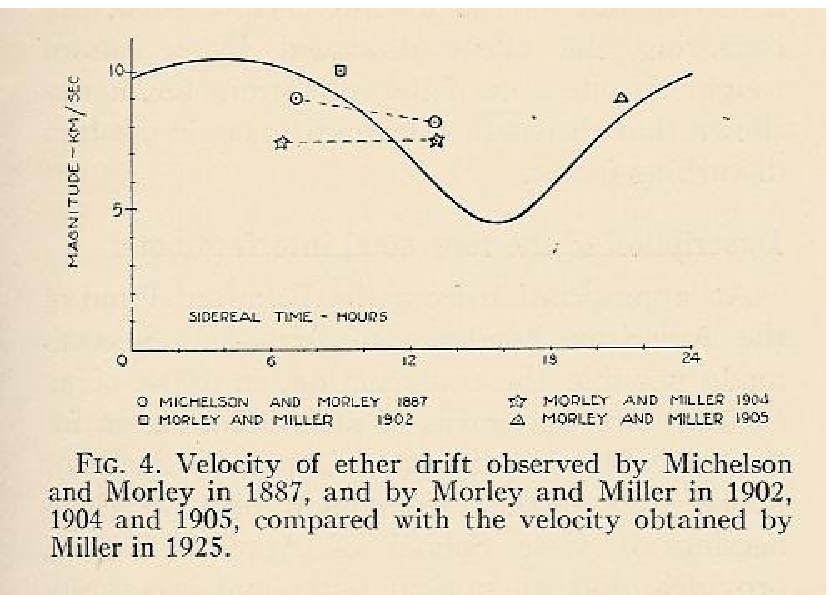}
\end{center}
\caption{\it The observable velocity Eq.(\ref{vobs2}) reported by
Miller \cite{miller} for various experiments.} \label{miller}
\end{figure}

Thus, no surprise that the resulting
${{|\Delta\bar{c}_\theta|}\over{c}} \sim \epsilon (v^2/c^2)$ was
much smaller than the classical expectation. For instance, in the
old experiments in air (at room temperature and atmospheric pressure
where $\epsilon\sim 2.8\cdot 10^{-4}$) a typical value was
${{|\Delta\bar{c}_\theta|_{\rm exp}}\over{c}} \sim 3\cdot 10^{-10}$.
By assuming the classical relations (\ref{a2class}),(\ref{vobs2}), this was originally interpreted as an observable velocity $v_{obs} \sim 7.3$ km/s but, by using Eq.(\ref{vobs}), this observable velocity would now correspond to a true kinematic velocity $v \sim 310$ km/s. Analogously, in the old experiment in gaseous helium (at room temperature and atmospheric pressure, where
$\epsilon\sim 3.3\cdot 10^{-5}$), a typical value was
${{|\Delta\bar{c}_\theta|_{\rm exp}}\over{c}} \sim 2.2\cdot
10^{-11}$. This was classically interpreted as an (observable) velocity of 2 km/s
but would now correspond to a true kinematical value $v \sim 240$ km/s.

Another observation concerns the time-dependence of the data and
the precise definition of $v$ and $\theta=0$ in the above
relations. Traditionally, for short-time observations of a few days,
where there can be no sizeable change in the orbital motion of the
Earth, the genuine signal for a preferred frame was consisting in
the regular modulations induced by the Earth rotation. Instead the
data had an irregular behavior indicating sizeably different
directions of the drift at the same hour on consecutive days. This
was a strong argument to interpret the small residuals as typical
instrumental artifacts. However, this conclusion derives from the traditional identification of the local velocity field
which describes the drift, say $v_\mu(t)$,
with the corresponding projection of the global Earth motion, say
$\tilde v_\mu(t)$. This identification is
equivalent to a form of regular, laminar flow where global and local
velocity fields coincide and, in principle, may be incorrect.

The model of ether drift adopted in refs.
\cite{plus,plus2,book,universe} starts from Maxwell's original
argument \cite{maxwell}.  After having considered all known
properties of light, he was driven to consider the idea of a
substratum: "...We are therefore obliged to suppose that the medium
through which light is propagated is something distinct from the
transparent media known to us...". He was calling this substratum
"ether" while, today, we prefer to call it "physical vacuum".
However, this is irrelevant. The essential point for the propagation
of light, e.g. inside an optical cavity, is that, differently from
the solid parts of the apparatus, this physical vacuum is not
totally entrained with the Earth motion. Therefore, to explain the
irregular character of the data, the original idea of
refs.\cite{chaos,physica} was to model this vacuum as a turbulent
fluid or, more precisely, as a fluid in the limit of zero viscosity
\footnote{The idea of the physical vacuum as an underlying
stochastic medium, similar to a turbulent fluid, is deeply rooted in basic foundational aspects
of both quantum physics and relativity. For instance, at the end of XIX century, the last model of the ether
was a fluid full of very small whirlpools (a
``vortex-sponge'') \cite{whittaker}. The hydrodynamics of this
medium was accounting for Maxwell's equations and thus 
providing a model of Lorentz symmetry as emerging from
a system whose elementary constituents are governed by Newtonian
dynamics. More recently, the turbulent-ether model has been
re-formulated by Troshkin \cite{troshkin} (see also \cite{puthoff}
and \cite{tsankov}) within the Navier-Stokes equation, 
by Saul \cite{saul} by starting from Boltzmann's transport
equation and  in \cite{pla12}  wihin Landau's hydrodynamics. 
The same picture of the
vacuum (or ether) as a turbulent fluid was Nelson's \cite{nelson}
starting point. In particular, the zero-viscosity limit gave him the
motivation to expect that ``the Brownian motion in the ether will
not be smooth'' and, therefore, to conceive the particular form of
kinematics at the base of his stochastic derivation of the
Schr\"odinger equation. A qualitatively similar picture is also
obtained by representing relativistic particle propagation from the
superposition, at short time scales, of non-relativistic
particle paths with different Newtonian mass \cite{kleinert}. In
this formulation, particles randomly propagate (as in a
Brownian motion) in an underlying granular medium which replaces the
trivial empty vacuum \cite{jizba}.}. Then, the simple picture of a
laminar flow is no more obvious due to the subtlety of the
infinite-Reynolds-number limit, see e.g. Sect. 41.5 in Vol.II of
Feynman's lectures \cite{feybook}. Namely, beside $v_\mu(t)=\tilde
v_\mu(t)$, there is also another solution where $v_\mu(t)$ is a
continuous, nowhere differentiable velocity field
\cite{onsager,eyink}. This leads to the idea of a signal with a
fundamental stochastic nature as when turbulence, at small scales,
becomes homogeneous and isotropic.  One should thus first analyze
the data for ${{\Delta\bar{c}_\theta(t)}\over{c}}$ and extract the
(2nd-harmonic) phase and amplitude $A_2(t)$ by concentrating on the
latter which is positive definite and remains non-zero under any
averaging procedure.

For a quantitative description, let us assume a set of kinematic
parameters $(V,\alpha,\gamma)$ for the Earth cosmic motion, a
latitude $\phi$ of the laboratory and a given sidereal time
$\tau=\omega_{\rm sid}t$ (with $\omega_{\rm sid}\sim
{{2\pi}\over{23^{h}56'}}$). Then $\tilde v(t)=V |\sin z(t)| $ is the
magnitude of the projection in the plane of the interferometer,
$\sin z(t)$ being defined by \cite{nassau}
\begin{equation} \label{nassau0}
       \cos z(t)= \sin\gamma\sin \phi + \cos\gamma
       \cos\phi \cos(\tau-\alpha)
\end{equation}
In this scheme, the amplitude $\tilde A_2(t)$ associated with the
global motion is \BE \label{smoothfinal} \tilde A_2(t) \sim {{2 D
}\over{\lambda}} \cdot \epsilon\cdot {{V^2 \sin^2 z(t)  } \over{c^2
}} \EE Although the local, irregular $v_\mu(t)$ is not a
differentiable function it could still be simulated in terms of
random Fourier series \cite{onsager,landau,fung}. This method was
adopted in refs.\cite{plus,plus2,book,universe} in a simplest
uniform-probability model, where the kinematic parameters of the
global $ \tilde v_\mu(t)$ are just used to fix the boundaries for
the local random $v_\mu(t)$. The essential ingredients are summarized in the Appendix.

We emphasize that the instantaneous, irregular $A_2(t)$ is very
different from the smooth $\tilde A_2(t)$. However, the relation
with the statistical average $\langle A_2(t) \rangle_{\rm stat}$ is
very simple  \BE \label{amplitude10001finalintro} \langle
A_2(t)\rangle_{\rm stat} \sim {{\pi^2 } \over{18 }}\cdot \tilde
A_2(t)\sim  {{2 D }\over{\lambda}} \cdot {{\pi^2 } \over{18 }}~
\epsilon ~\frac{V^2 \sin^2 z(t)}{c^2} \EE Furthermore, by using
Eq.(\ref{nassau0}), if the amplitude is measured at various
sidereal times, one can also get information on the angular
parameters $\alpha$ and $\gamma$.

Altogether, those old measurements
${{|\Delta\bar{c}_\theta|_{\rm exp}}\over{c}} \sim 3\cdot 10^{-10}$
and ${{|\Delta\bar{c}_\theta|_{\rm exp}}\over{c}} \sim 2.2\cdot
10^{-11}$, respectively for air or gaseous helium at atmospheric
pressure, can thus be interpreted in three different ways: a) as 7.3 and
2 km/s, in a classical picture b) as 310 and 240 km/s, in a modern
scheme and in a smooth picture of the drift c) as 418 and 324 km/s,
in a modern scheme but now allowing for irregular fluctuations of
the signal. In this last case, in fact, by replacing Eq.(\ref{smoothfinal}) with Eq.(\ref{amplitude10001finalintro}), from the same data one would now obtain kinematical velocities which are larger by a factor $\sqrt{18/{\pi^2}} \sim 1.35$. In this third interpretation, the average of the two
values agrees very well with the CMB velocity of 370 km/s. The
comparison with all classical experiments is shown in Table
\ref{summary}.

\begin{table}[h]
\caption{\it The average 2nd-harmonic amplitudes of classical
ether-drift experiments. These were extracted from the original
papers by averaging the amplitudes of the individual observations
and assuming the direction of the local drift to be completely
random (i.e. no vector averaging of different sessions). These
experimental values are then compared with the full statistical
average Eq.(\ref{amplitude10001finalintro}) for a projection 250
km/s $\lesssim V| \sin z(t)| \lesssim$ 370 km/s of the Earth motion in the CMB
and refractivities $\epsilon=2.8\cdot10^{-4}$ for air and
$\epsilon=3.3\cdot10^{-5}$ for gaseous helium. The experimental
value for the Morley-Miller experiment is taken from the observed
velocities reported in Miller's Figure 4, here our Fig.\ref{miller}.
The experimental value for the Michelson-Pease-Pearson experiment
refers to the only known session for which the fringe shifts are
reported explicitly \cite{pease} and where the optical path was
still fifty-five feet. The symbol $\pm ....$ means that the
experimental uncertainty cannot be determined from the available
informations. The table is taken from ref.\cite{universe}.}
\begin{tabular}{cllll}
\hline Experiment &gas
&~~~~$A^{\rm EXP}_2$ &~~~ ${{2D}\over{\lambda}}$~~~~& ~~~ $\langle A_2(t)\rangle_{\rm stat} $   \\
\hline
Michelson(1881)               & air     &$ (7.8 \pm....)\cdot10^{-3}$     &~~~$4\cdot 10^6  $   &$(0.7 \pm 0.2)\cdot 10^{-3}$  \\
Michelson-Morley(1887)   & air & $(1.6 \pm 0.6)\cdot 10^{-2 }$&~~~$4\cdot 10^7$ & $(0.7 \pm 0.2)\cdot 10^{-2}$ \\
Morley-Miller(1902-1905)   & air & $(4.0 \pm 2.0)\cdot 10^{-2 }$&~~~$1.12\cdot 10^8$ &$ (2.0 \pm 0.7) \cdot10^{-2}$\\
Miller(1921-1926)  & air& $(4.4 \pm 2.2)\cdot 10^{-2 }$ & ~~  $1.12\cdot 10^8$ &$(2.0 \pm 0.7) \cdot10^{-2} $ \\
Tomaschek (1924) & air & $(1.0\pm 0.6)\cdot 10^{-2 }  $ &~~~$3\cdot 10^7$& $ (0.5 \pm 0.2) \cdot10^{-2} $\\
Kennedy(1926)  & helium & ~~~$<0.002$&~~~$7 \cdot 10^6$&$ (1.4 \pm 0.5)\cdot10^{-4}  $\\
Illingworth(1927) & helium & $ (2.2 \pm 1.7)\cdot 10^{-4}  $  &~~~$7 \cdot 10^6$ &$ (1.4 \pm 0.5)\cdot10^{-4}$ \\
Piccard-Stahel(1928)          &air & $(2.8 \pm 1.5)\cdot10^{-3}$  &~~~$1.28 \cdot 10^7$& $(2.2 \pm 0.8)\cdot10^{-3}$\\
Mich.-Pease-Pearson(1929) & air& $(0.6 \pm...)\cdot10^{-2}$  &~~~$5.8  \cdot 10^7$& $(1.0 \pm 0.4)\cdot10^{-2}$\\
Joos(1930)  &helium&$(1.4 \pm 0.8)\cdot 10^{-3 }   $  & ~~ $7.5 \cdot 10^7$&$(1.5 \pm 0.6)\cdot10^{-3}$\\
\hline
\end{tabular}
\label{summary}
\end{table}

Notice the substantial difference with the analogous summary Table I
of ref.\cite{shankland} where those authors were comparing with the
much larger classical amplitudes  Eq.(\ref{a2class}) and emphasizing
the much smaller magnitude of the experimental fringes. Here, is
just the opposite. In fact, our theoretical statistical averages are
often {\it smaller} than the experimental results indicating, most
likely, the presence of systematic effects in the measurements.

At the same time, by adopting
Eq.(\ref{amplitude10001finalintro}), from the experiments in air we find
$\tilde v_{\rm air} \sim 418 \pm 62 $ km/s and from the two experiments
in gaseous helium $\tilde v_{\rm helium} \sim 323 \pm 70 $ km/s,
with a global average $\langle\tilde v \rangle\sim 376 \pm 46 $ km/s
which agrees well with the 370 km/s from the CMB
observations.  Even more, from the two most precise
experiments of Piccard-Stahel (the measurements performed on ground in Bruxelles and at Mt. Rigi in
Switzerland) \footnote{In ref.\cite{book} a numerical simulation of
the Piccard-Stahel experiment \cite{piccard3} is reported, for both
the individual sets of 10 rotations of the interferometer and the
experimental sessions (12 sets, each set consisting of 10
rotations). Our analysis confirms their idea that the optical path
was much shorter than the instruments in United States but their
measurements were more precise because spurious disturbances were
less important.} and those made in Jena by Georg Joos  \footnote{Joos' optical system
was enclosed in a hermetic housing and, as reported by Miller
\cite{miller,miller34}, it was traditionally believed that his
measurements were performed in a partial vacuum. In his article,
however, Joos is not clear on this particular aspect. Only when
describing his device for electromagnetic fine movements of the
mirrors, he refers to the condition of an evacuated apparatus
\cite{joos}. Instead, Swenson \cite{swensonbook,loyd2} declares that
Joos' fringe shifts were finally recorded with optical paths placed
in a helium bath. Therefore, we have followed Swenson's explicit
statements and assumed the presence of gaseous helium at atmospheric
pressure.} we find, in our stochastic scheme, two determinations,
$\tilde v= 360^{+85}_{-110} $ km/s and $\tilde v= 305^{+85}_{-100} $
km/s respectively, whose average $\langle \tilde v\rangle \sim
332^{+60}_{-80} $ km/s reproduces to high accuracy the projection of
the CMB velocity at a typical Central-Europe latitude.
Finally, by using Eq.(\ref{nassau0}) and
fitting the amplitudes obtained from Joos' observations
(data collected at steps of 1 hour to cover the sidereal day) one finds \cite{plus,book}
$\alpha({\rm fit-Joos})= (168 \pm 30)$ degrees and
$\gamma({\rm fit-Joos})= (-13 \pm 14)$
degrees which are consistent with the present values $\alpha({\rm
CMB}) \sim$ 168 degrees and $\gamma({\rm CMB}) \sim -$7 degrees.

As it often happens, symmetry arguments can successfully describe a
phenomenon regardless of the physical mechanisms behind it. The same
is true here with our relation
${{|\Delta\bar{c}_\theta|}\over{c}}\sim \epsilon (v^2/c^2)$. It
gives a consistent description of the data but does not explain the ultimate
origin of the tiny observed anisotropy in the
gaseous systems. For instance, as a first mechanism, we considered
the possibility of different polarizations in different
directions in the dielectric, depending on its state of motion.
However, if this works in weakly bound gaseous matter, the same
mechanism should also work in a strongly bound solid dielectric,
where the refractivity is $({\cal N}_{\rm solid} -1)= O(1)$, and
thus produce a much larger ${{|\Delta\bar{c}_\theta|}\over{c}}\sim
({\cal N}_{\rm solid} -1) (v^2/c^2)\sim 10^{-6} $. This is in
contrast with the Shamir-Fox \cite{fox} experiment in perspex where
the observed value was smaller by orders of magnitude. We have thus
re-considered \cite{epl,plus2,book} the traditional thermal
interpretation \cite{joos2,shankland} of the observed residuals. The
idea was that, in a weakly bound system as a gas, a small
temperature difference $\Delta T^{\rm gas}(\theta)$, of a
millikelvin or so, in the air of the two optical arms could produce
a difference in the refractive index and a light anisotropy
proportional to $\epsilon_{\rm gas} \Delta T^{\rm gas}(\theta)/T$,
where $T\sim$ 300 K is the temperature of the laboratory. Miller was
aware of this potentially large effect \cite{miller,miller34} and
objected that casual changes of the ambiance temperature would
largely cancel when averaging over many measurements. Only
temperature effects with a definite angular periodicity would survive.
The overall consistency, in our scheme, of different experiments
would now indicate that such $\Delta T^{\rm gas}(\theta)$ must have
a {\it non-local} origin. As anticipated, this could be due to the
non-zero momentum flow Eq.(\ref{nonzeroP_i}). Or it could reflect
the interactions with the background radiation which transfer a part
of $\Delta T^{\rm CMB}(\theta)$ in Eq.(\ref{CBR}) and thus bring the
gas out of equilibrium. Those old estimates were, however,
slightly too large. In fact, in the ideal-gas approximation,
from the Lorentz-Lorenz equation for the molecular
polarizability, one actually finds
\cite{plus2,book,universe} $\Delta T^{\rm gas}(\theta)= (0.2-0.3)$ mK \footnote{Interestingly, after a century from those old
experiments, in a room-temperature ambiance, the fraction of
millikelvin is still state of the art when measuring temperature
differences, see \cite{farkas,zhaoa,trusov}. This supports our idea
that  $\Delta T^{\rm gas}(\theta)$ is a non-local effect which
places a fundamental limit.}. For the CMB case, this would mean that
the interactions of the gas with the CMB photons are so weak that,
on average, the induced temperature differences in the optical paths
were only 1/10 of the $\Delta T^{\rm CMB}(\theta)$ in
Eq.(\ref{CBR}). Nevertheless, whatever its precise origin, this
typical magnitude can help intuition. In fact, it can explain the
{\it quantitative} reduction of the effect in the vacuum limit where
$\epsilon_{\rm gas} \to 0$ and the {\it qualitative} difference with
solid dielectrics where such small temperature differences cannot
produce any appreciable deviation from isotropy in the rest frame of
the medium.

Now, admittedly,  the idea that small modifications of the gaseous
matter, produced by the  tiny CMB temperature variations, can be
detected  by precise optical measurements in a laboratory, while
certainly unconventional, has not the same implications of a genuine
preferred-frame effect due to the vacuum structure. Still, this
thermal explanation of the small residuals in gases has an important
predictive power. In fact, it implies that if a tiny, but non-zero,
fundamental signal were definitely detected in vacuum then, with
very precise measurements, the same universal signal should also
show up in a solid dielectric where temperature differences of a
fraction of millikelvin become irrelevant. Detecting such
`non-thermal' light anisotropy, for the same cosmic motion indicated
by the CMB observations, would finally confirm the idea of
$\Sigma-$frame assumed in refs.
\cite{scarani,cocciaro,salart,bancal,cocciaro2}.

\subsection{The modern experiments in vacuum and solid dielectrics}

\begin{figure}[h]
\begin{center}
\includegraphics[width=7.0 cm]{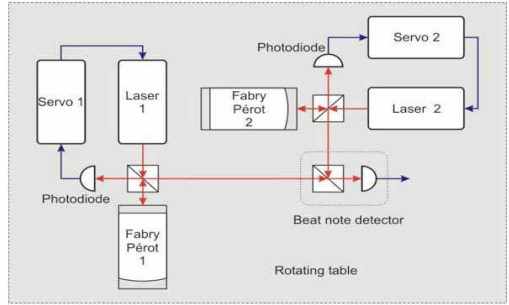}
\caption{\it The scheme of a modern ether-drift experiment. The
light frequencies are first stabilized by coupling the lasers to
Fabry-Perot optical resonators. The frequencies $\nu_1$ and $\nu_2$
of the resonators are then compared in the beat note detector which
provides the frequency shift $\Delta \nu(\theta)=\nu_1(\pi/2 +
\theta) -\nu_2(\theta)$. For a review, see e.g. \cite{applied}. } \label{Fig.apparatus}
\end{center}
\end{figure}

This expectation of a `non-thermal' light anisotropy, which could be
detected in vacuum and in solid dielectrics, was then compared with
the modern experiments where ${{\Delta\bar{c}_\theta}\over{c}}\sim
{{\Delta\nu(\theta)}\over{\nu_0}}$ is now extracted from the
frequency shift of two optical resonators see
Fig.\ref{Fig.apparatus}.

\begin{figure}[h]
\begin{center}
\includegraphics[width=7.0 cm]{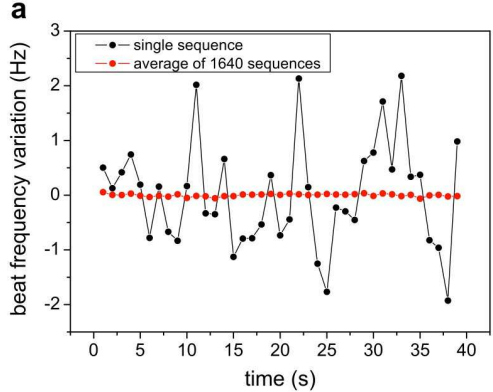}
\end{center}
\caption{\it The experimental frequency shift reported in Fig.9(a)
of ref.\cite{crossed} (courtesy Optics Communications). The black
dots give the instantaneous signal, the red dots give the signal
averaged over 1640 sequences. For a laser frequency $\nu_0=2.8\cdot
10^{14}$ Hz a $\Delta \nu=\pm 1$ Hz corresponds to a fractional
value $\Delta \nu/\nu_0$ of about $\pm 3.5 \cdot 10^{-15}$.}
\label{Figcrossed}
\end{figure}

By starting with vacuum resonators, after averaging many
observations, the present limit is a residual $\langle
{{\Delta\bar{c}_\theta}\over{c}}\rangle =10^{-18}-10^{-19}$.
However, this just reflects the very irregular nature of the signal
because its typical {\it instantaneous} magnitude
${{|\Delta\bar{c}_\theta(t)|_v}\over{c}}\sim 10^{-15}$ is about 1000
times larger, see Fig.\ref{Figcrossed}. This $10^{-15}$ signal is
found with vacuum resonators
\cite{mueller2003}$-$\cite{schiller2015} made of different
materials, operating at room temperature and/or in the cryogenic
regime. As such, it cannot be interpreted as a spurious effect, e.g.
thermal noise \cite{numata}.

\begin{figure}[h]
\begin{center}
\includegraphics[scale=0.30]{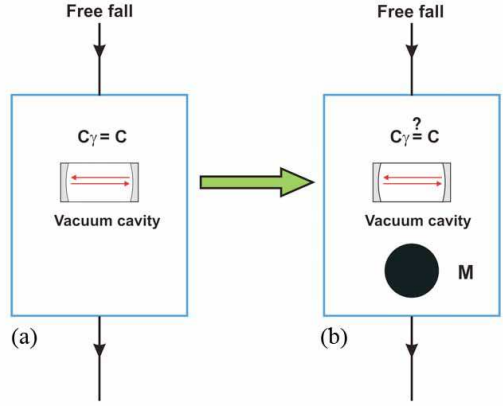}
\end{center}
\caption{ {\it A heavy mass $M$ is carried on board of a freely-falling system, case (b). With
respect to the ideal case (a), the mass $M$ modifies the local
space-time units and could introduce a vacuum refractive
index ${\cal N}_v\neq 1$ so that now $c_\gamma \neq c$. With a preferred frame, one would then expect
off-diagonal elements $g_{0i}\sim2({\cal N}_v -1) (v_i/c)$ in the effective metric which describes light propagation for the (b) reference system.} }
\label{freefall}
\end{figure}

In the same model discussed for the classical experiments, we are
thus lead to the concept of a refractive index ${\cal N}_v$ for the
vacuum or, more precisely, for the physical vacuum which is
established in an apparatus placed on the Earth surface. This ${\cal
N}_v$ should differ from unity at the $10^{-9}$ level, in order to
give $ {{|\Delta\bar{c}_\theta(t)|_v}\over{c}}\sim ({\cal N}_v
-1)~(v^2(t)/c^2)~ \sim 10^{-15}$, and thus would fit with
ref.\cite{gerg} where, for an apparatus placed on the Earth surface,
a vacuum refractivity $\epsilon_v\sim (2G_NM/c^2R) \sim 1.4\cdot
10^{-9}$ was considered, $G_N$ being the Newton constant and $M$ and
$R$ the mass and radius of the Earth. The idea is that, if the
curvature observed in a gravitational field reflects local
deformations of the physical space-time units and of the velocity of
light  \cite{broekaert}, for an apparatus on the Earth surface, there could be a tiny
difference with that ideal free-fall environment which, in the presence of gravitational effects, is always
assumed to define operationally the limit where the velocity of light in vacuum
$c_\gamma$ coincides with the
parameter $c$ of Lorentz
transformations. This would
reflect the physical difference which, indeed, exists \cite{gerg} between an observer in
a true free-falling elevator and the modified situation of an
observer which is in free fall in the same external potential but is
now carrying on board a heavy mass $M$, see Fig.\ref{freefall}.

Therefore, if $\delta U$ is the extra Newtonian potential
produced by the heavy mass $M$ at the experimental setup, the vacuum
refractivity, for system {\bf (b)}, can be expressed as
\begin{equation} \label{refractive}
\epsilon_v={\cal N}_v - 1 \sim {{\chi}\over{2}}~\left({{2|\delta
U|}\over{c^2}}\right)
\end{equation}
In General Relativity one assumes $\chi=0$ while the two non-zero
values, $\chi=$ 1 or 2, account for the two alternatives
traditionally reported in the literature for the effective
refractive index in a gravitational potential. For $\chi=2$, the
resulting refractivity is the same reported by Eddington
\cite{eddington} to explain in flat space the observed deflection of
light in a gravitational field. A difference is found with Landau's
and Lifshitz' textbook \cite{landaufield} where the vacuum
refractive index entering the constitutive relations is instead
defined as ${\cal N}_v \sim 1+{{|\delta U|}\over{c^2}}$. We address
to Broekaert's article \cite{broekaert}, and in particular to his
footnote $^{3}$, for a more detailed discussion of the two choices
of $\chi$.

In our case, of an observer on the Earth surface, by introducing the Newton constant, the radius $R$ and
the mass $M$ of the Earth, so that $\delta U={{G_N M}\over{R}}$, we
can express the refractivity as
\begin{equation} \label{refractive2}
\epsilon_v \sim {{\chi}\over{2}}~1.4\cdot 10^{-9} \end{equation}
Addressing to ref.\cite{universe}, here we just
report the comparison with ref.\cite{schiller2015} which, at
present, is the most precise experiment in vacuum. We compared with
the average instantaneous variation of the
frequency shift over 1 second, see their Fig.3, bottom part. This is
defined by the Root Square of the Allan Variance (RAV) \footnote{The
RAV gives the variation of a function $f=f(t)$
sampled over steps of time $\tau$. By
defining \BE {\overline f}(t_i;\tau)={{1}\over{\tau }}\int^{t_i+\tau
}_{t_i}dt~f(t)\equiv {\overline f}_i \nonumber\EE one generates a
$\tau-$dependent distribution of ${\overline f}_i$ values. In a
large time interval $\Lambda= M\tau$, the RAV is then defined as \BE
\sigma_A(f,\tau)= \sqrt{\sigma^2_A(f,\tau) } \nonumber\EE where \BE
\sigma^2_A(f,\tau)= {{1}\over{2(M-1) }}\sum^{M-1}_{i=1}
\left({\overline f}_i-{\overline f}_{i+1} \right)^2 \nonumber\EE The
integration time $\tau$ is given in seconds and the factor of 2 is
introduced to obtain the same standard variance for uncorrelated
data as for a white-noise signal with uniform spectral amplitude at
all frequencies.}(for $\tau_0\sim $ 1 second)
\begin{equation} \left[ \sigma_A(\Delta \nu,\tau_0) \right]_{\rm
exp}\sim 0.24~ {\rm Hz}
\end{equation}
or, in units of the reference frequency $\nu_0=2.8\cdot 10^{14}$ Hz
(for $\tau_0\sim $ 1 second)
\begin{equation} \label{ravschiller2015}\left[\sigma_A(\frac{ \Delta \nu}{\nu_0},\tau_0) \right]_{\rm
exp}~\sim 8.5 \cdot 10^{-16} ~~~~~~~~~~~~~~~~~~{\rm ref.
\cite{schiller2015}}
\end{equation}
As discussed in ref.\cite{universe}, our instantaneous, stochastic
signal is, to very good approximation, a pure white noise for which
the RAV coincides with the standard variance. At the same time, for
a very irregular signal where $\langle\Delta \nu\rangle=0$ the
standard variance $\sigma(\Delta \nu)$ coincides with the average
magnitude $\langle | \Delta \nu |\rangle$. Therefore, since in our
stochastic model, the average magnitude of the dimensionless
frequency shift ${{\Delta\nu(\theta)}\over{\nu_0}}\sim
{{\Delta\bar{c}_\theta}\over{c}} $ is given in
Eq.(\ref{amplitude10001finalintro}), we find (for $\tau_0\sim $ 1
second)
\begin{equation} \label{deltanuth}
\left[\sigma_A(\frac{ \Delta \nu}{\nu_0},\tau_0) \right]_{\rm
theor}\sim \left[ \frac{ \langle |\Delta \nu|\rangle
}{\nu_0}\right]_{\rm theor}~\sim {{\pi^2 } \over{18 }} \cdot
\epsilon_v \cdot{{V^2  } \over{c^2 }} \sin^2 z(t)
\end{equation}
In this way, by replacing Eq.(\ref{refractive2}), and for a
projection 250 km/s $\lesssim V\sin z(t)\lesssim$ 370 Km/s, for $\tau_0\sim
$ 1 second, our prediction for the RAV can finally be expressed as
\begin{equation} \label{allanth}\left[\sigma_A(\frac{ \Delta \nu}{\nu_0},\tau_0) \right]_{\rm
theor}~\sim {{\chi}\over{2}}\cdot(8.5 \pm 3.5) \cdot 10^{-16}
\end{equation}

By comparing with Eq.(\ref{ravschiller2015}), we see that the data
definitely favor $\chi=2$, which is the only free parameter of our
scheme. Also, the very good agreement with our simulated value
indicates that, at least for an integration time of 1 second, the
corrections to our model should be negligible \footnote{Numerical simulations indicate that our vacuum signal has the same characteristics of
a universal white noise. Thus, strictly speaking, it should be compared with the
frequency shift of two optical resonators at the largest integration
time $\tau_0$ where the pure white-noise branch is as
small as possible but other types of noise are not yet important. In the experiments
we are presently considering this $\tau_0$ is typically 1 second. However, in principle, $\tau_0$
could also be considerably larger than 1 second as, for instance, in
the cryogenic experiment of ref.\cite{mueller2003}. There, the RAV
at 1 second was about 10 times larger than the range
Eq.(\ref{allanth}) but, in the quiet phases between two refills of the
refrigerator, $\sigma_A(\Delta \nu/\nu_0,\tau)$ was monotonically
following the white-noise trend $\tau^{-1/2}$ up to $\tau_0\sim 240$ seconds where it
reached its minimum value $\sigma_A(\Delta \nu/\nu_0,\tau_0)\sim
5.3\cdot 10^{-16}$. Remarkably, for $\chi=2$, this is still consistent with the theoretical range
Eq.(\ref{allanth}). }.

Let us now compare with the modern experiments in solid dielectrics,
in particular with the very precise ref.\cite{nagelnature}. This is a cryogenic
experiment, with microwaves of 12.97 GHz, where almost all
electromagnetic energy propagates in a medium, sapphire, with
refractive index of about 3 (at microwave frequencies). As
anticipated, with a thermal interpretation of the residuals in
gaseous media, we expect that a fundamental $10^{-15}$ vacuum
anisotropy could also become visible here.

Following refs.\cite{plus2,book,universe}, we first observe that for
${\cal N}_v=1 +\epsilon_v$ there will be a very tiny difference
between the refractive index defined relatively to the ideal vacuum
value $c$ and the refractive index relatively to the physical
isotropic vacuum value $c/{\cal N}_v$ measured on the Earth surface.
The relative difference between these two definitions is
proportional to $\epsilon_v \lesssim 10^{-9} $ and, for all
practical purposes, can be ignored. All materials would now exhibit,
however, the same background vacuum anisotropy. To this end, let us
replace the average isotropic value \BE {{c}\over { {\cal N}_{\rm
solid}}} \to {{c}\over { {\cal N}_v {\cal N}_{\rm solid} }}\EE and
then  use Eq.(\ref{nbartheta}) to replace ${\cal N}_v$ in the
denominator with its $\theta-$dependent value \BE \bar {\cal
N}_v(\theta) \sim  1+ \epsilon_v\beta^2(1+\cos^2\theta)\EE This is
equivalent to define a $\theta-$dependent refractive index for the
solid dielectric
\begin{eqnarray}\label{6introsolid}
        {{  \bar{\cal N}_{\rm solid}(\theta)}\over { {\cal N}_{\rm solid}}} \sim  1+\epsilon_v \beta^2(1
        +
       \cos^2\theta)
\end{eqnarray}
so that
\begin{equation}
\label{refractivetheta1} \left[ {\bar c_\gamma (\theta)}
\right]_{\rm solid}={{c}\over{\bar{\cal N}_{\rm
solid}(\theta)}} \sim {{c}\over { {\cal N}_{\rm solid}}} \left[ 1-
\epsilon_v \beta^2 (1 +
       \cos^2\theta)\right]
\end{equation}
with an anisotropy
\begin{equation}
\label{anysolid} {{ \left[\Delta\bar{c}_\theta\right]_{\rm solid}}
\over {\left[ c/ {\cal N}_{\rm solid}\right] }} \sim \epsilon_v
\beta^2 \cos2\theta
\end{equation}
In this way, a genuine vacuum effect, if there, could also be
detected in a solid dielectric.

\begin{figure}[h]
\begin{center}
\includegraphics[scale=0.25]{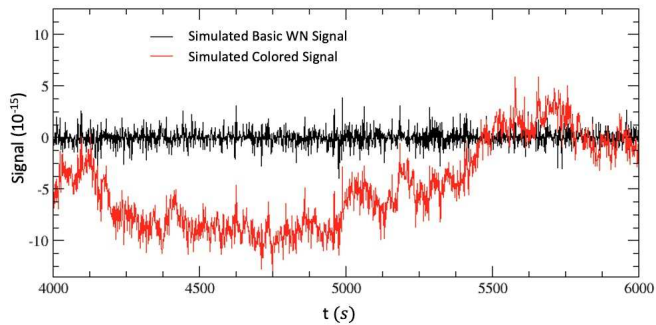}
\end{center}
\caption{\it We report two typical sets of 2000 seconds for our
basic white-noise (WN) signal and its colored version obtained by
Fourier transforming the spectral amplitude of
ref.\cite{nagelnature}. The boundaries of the random velocity
components Eqs.(\ref{vx}) and (\ref{vy}) were defined by
Eq.(\ref{isot}) by plugging in Eq.(\ref{projection}) the CMB
kinematical parameters, for a sidereal time $t=4000-6000$
seconds and for the latitude of Berlin-Duesseldorf, see the
Appendix. The figure is taken from ref.\cite{universe}.}
\label{colored}
\end{figure}

In ref.\cite{universe}, a detailed comparison with
\cite{nagelnature} was presented. First, from Figure 3(c) of
\cite{nagelnature}, it was seen that the spectral amplitude of this
particular apparatus  becomes flat at frequencies $\omega \ge 0.5$
Hz indicating that the white-noise branch of the signal reaches its
minimum value for an integration time $\tau_0\sim$ 1 second (at
which the other spurious disturbances are still negligible). These
data for the spectral amplitude were then fitted to an analytic,
power-law form to describe the lower-frequency part 0.001 Hz $\leq
\omega \leq 0.5$ Hz which reflects apparatus-dependent
disturbances. This fitted spectrum was then used to generate
a signal by Fourier transform. Finally, very long sequences of this
signal were stored to produce ``colored'' version of our basic
white-noise signal.

\begin{figure}[h]
\begin{center}
\includegraphics[scale=0.25]{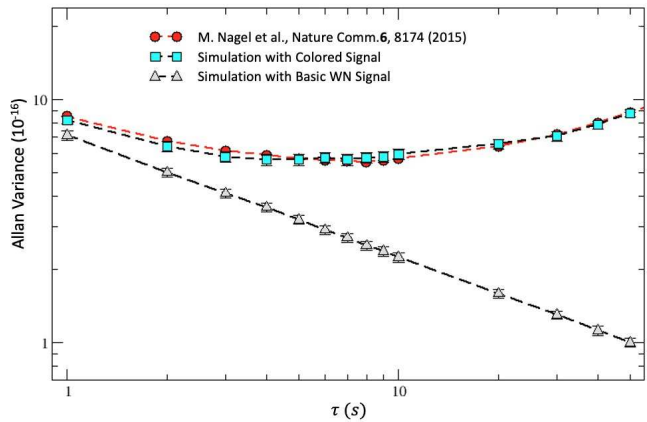}
\end{center}
\caption{\it We report the Allan variance for the fractional
frequency shift obtained from many simulations of sequences of 2000
seconds for our basic white-noise (WN) signal and for its colored
version, see Fig.\ref{colored}. The direct experimental results of
ref.\cite{nagelnature}, for the non-rotating setup, are also shown.
The figure is taken from ref.\cite{universe}.} \label{allan}
\end{figure}

To get a qualitative impression of the effect, we report in
Fig.\ref{colored} a sequence of our basic white-noise signal and a
sequence of its colored version. By averaging over many 2000-second
sequences of this type, the corresponding RAV's for the two signals
are then reported in Fig.\ref{allan}. The experimental RAV extracted
from Figure 3(b) of ref.\cite{nagelnature} is also reported (for the
non-rotating setup). At this stage, the agreement of our simulated,
colored signal with the experimental data remains satisfactory only
up $\tau=$ 50 seconds. Reproducing the signal at larger $\tau$'s
will require further efforts but this is not relevant here, our
scope being just to understand the modifications of our stochastic
signal near the 1-second scale.

As one can check from Fig.3(b) of ref.\cite{nagelnature}, the value
of the experimental RAV for the fractional frequency shift (at
$\tau_0=1$ second) is \BE \label{ravnagelnature}\sigma_A({{\Delta
\nu }\over{\nu_0}},\tau_0)_{\rm exp}\sim  8.5 \cdot 10^{-16}
~~~~~~~~~~~~~~~~~~{\rm ref. \cite{nagelnature}}\EE  This is
precisely the same value Eq.(\ref{ravschiller2015}) that we
extracted from  ref.\cite{schiller2015} after normalizing their
experimental result $\sigma_A(\Delta \nu,\tau_0)_{\rm exp}\sim $
0.24 Hz  to their laser frequency $\nu_0=2.8 \cdot 10^{14}$ Hz. At
the same time, it also agrees with our Eq.(\ref{allanth})
$\sigma_A({{\Delta \nu }\over{\nu_0}},\tau_0)_{\rm theor}=(8.5 \pm
3.5)\cdot 10^{-16}$ for $\chi=2$. Therefore this beautiful
agreement, between ref.\cite{schiller2015} (a vacuum experiment at
room temperature) and ref.\cite{nagelnature} (a cryogenic experiment
in a solid dielectric), on the one hand, and with our
Eq.(\ref{allanth}), on the other hand, confirms our interpretation
of the experiments in terms of a stochastic signal associated with the
Earth cosmic motion within the CMB.

Two ultimate experimental checks still remain. First, one should try
to detect our predicted, daily variations Eq.(\ref{allanth}) in the
range $(5-12) \cdot 10^{-16}$ corresponding to 250 km/s $\lesssim V\sin z(t)\lesssim$ 370 Km/s.
Due to the excellent systematics,
these should remain visible with both experimental setups. Second,
one more complementary test should be performed by placing the
vacuum (or solid dielectric) optical cavities on board of a
satellite, as in the OPTIS proposal \cite{optis}. In this ideal
free-fall environment, as in panel (a) of our Fig.\ref{freefall},
the typical instantaneous frequency shift should be much smaller (by
orders of magnitude) than the corresponding $10^{-15}$ value
measured with the same interferometers on the Earth surface.

\section{Summary and outlook}

In this paper, we have considered one of the most controversial
aspects of Quantum Mechanics, namely the apparent violation of
Einstein locality and the conflict with (Einstein) relativity. Since
the original paper by Einstein-Podolski-Rosen (EPR) \cite{EPR} and
through the work of Bell \cite{Bell}, many authors have thus arrived
to the conclusion that, to dispose of the causality paradox in a
realistic interpretation of the theory, it may be natural to
introduce a special frame of reference $\Sigma$ for relativity.
Then, one can consider the idea that some `Quantum Information'
propagates at a vastly superluminal speed $v_{QI} \gg c$, with
standard Quantum Mechanics corresponding to the $v_{QI}\to \infty$
limit. In this way, by comparing with experiments
\cite{scarani,cocciaro,salart,bancal,cocciaro2} one finds the lower
bounds $v_{QI}> 10^4 -  10^6 c$ if the preferred $\Sigma-$frame is
identified with the reference system where the Cosmic Microwave
Background (CMB) is seen isotropic, namely that particular system
where the observed CMB Kinematic Dipole \cite{yoon} vanishes
exactly.

A frequent objection to this idea of a preferred frame is that,
after all, Quantum Mechanics is not a fundamental description of the
world. One should instead start from a fundamental QFT which
incorporates the locality requirement. For this reason we have tried
to understand if, in the perspective of an underlying, fundamental
QFT, there could be a missing logical step which prevents to deduce
that Einstein Special Relativity, with no preferred frame, is the
{\it physically realized} version of relativity. Einstein Relativity
is always assumed when computing S-matrix elements for elementary
particle processes but what one is actually using is the machinery
of Lorentz transformations whose first, complete derivation dates
back to Larmor and Lorentz who were assuming the existence of a
fundamental state of rest (the ether).

In particular, in our opinion, an element missed so far derives from
the present view of the lowest energy state in elementary particle
theory which is called "vacuum" but should actually be thought as a
condensate of elementary quanta. So far, one has always been
imposing the traditional constraint that, as far as local operators
are concerned, the only possible vacuum expectation values are those
which transform as world scalars under the Lorentz group.

But, as discussed in Sect.2, this does {\it not} imply the vacuum to be a Lorentz-invariant state. This would
rather require the vacuum to be annihilated by the generators of the
Lorentz boosts. In four space-time dimensions, this requirement is only fulfilled for
the free-field case which, by definition, has a unique vacuum and
where the simplest prescription of the Wick, normal ordering allows
for a consistent representation of the Poincar\'e algebra. Instead,
in the interacting theory and in the presence of Spontaneous Symmetry Breaking, one can meaningfully argue 
that the {\it physically realized} form of relativity contains a
preferred reference system $\Sigma$.

However, since our arguments have not the status of a theorem, to decide
if $\Sigma$ really exists we have then looked for definite
experimental indications from the ether-drift experiments by
summarizing in Sect.3 the extensive work of
refs.\cite{plus}-\cite{epl}, where all data from Michelson-Morley to
the present experiments with optical resonators were considered.
Ours is not the only possible scheme to analyze the data but, still,
in this theoretical framework, which assumes the validity of Lorentz
transformations and allows for irregular fluctuations of the signal,
the long sought $\Sigma-$frame tight to the CMB is naturally
emerging.

Finally, before closing our paper, we will return to our starting
point: the idea that  eventually the non-locality
of Quantum Mechanics could be understood as the consequence of some `Quantum Information' which
propagates at a vastly superluminal speed $v_{QI}\gg c$. This was, after all, Bell's
conviction, namely that his result combined with the EPR argument, implies nonlocal
physical effects, and not just correlations between distant events \cite{Bricmont}. More
specifically if, as we have argued, the physical vacuum is really
the ultimate origin of the $\Sigma-$frame, the hypothetical
superluminal effects could be hidden somewhere in the physical
structure of the condensed vacuum. To exploit this possibility, we will assume
that this physical vacuum, however different from ordinary matter,
is nevertheless a medium with a certain degree of substantiality. As such,
it should exhibit density fluctuations. In this case,
these density fluctuations would propagate with a speed
$c_s \gg c$. We believe that,  in the present context, this can be a relevant issue,
even without a definite model where the previous $v_{QI}\gg c$ is directly related to a $c_s\gg c$.

To this end, we first recall that, as anticipated in the Introduction,``it is an
open question whether $c_s/c$ remains less than unity when
non-electromagnetic forces are taken into
account''\cite{Weinbergsound}. This is why superluminal
sound has been meaningfully considered by several authors see e.g.
refs.\cite{KP,BR1,R,BR2,kp}. The point is that the sought non-local effect may derive from two different space-time
regions. The first region is universal and is associated with the
localization of the interacting particles, i.e. their Compton
wavelength. This type of effects remain confined to microscopic distances. The
second region, on the other hand, depends on the basic interaction
which, dealing with non-electromagnetic interactions, as in our case
of a hard-sphere cutoff $\Phi^4$ theory, could be instantaneous.
Therefore, if each successive event leads to a small violation of
causality, with a sufficiently long chain of scattering events, the
effect could be amplified to macroscopic distances. Notice that we
are not speaking of scattering events with single-particle
propagation over large distances. In Bose condensates, each
particle moves slowly back and forth of a very small amount and it only
scatters with those particles which are immediately nearby. It is
the coherent effect of these local scattering processes which
propagates at much higher speed \cite{paulsound} and could produce, in principle, a faster-than-light
sound wave.

With this premise, in a pure hydrodynamic description, valid over length scales much larger than
the mean free path of the elementary constituents,  an argument for
superluminal sound could be the following. Let us consider the basic
relation \BE P= -{\cal E} + n \frac{d{\cal E}}{dn} \EE which relates
the pressure $P$ and the energy density ${\cal E}$ in a medium of
density $n$. By expanding the energy density around some given value $n=n_0$, we find
\BE {\cal E} \sim  {\cal E}(n_0) + A\cdot (n-n_0) +\half B\cdot (n-n_0)^2 \EE
\BE P \sim  - {\cal E}(n_0)   + A\cdot  n_0 +\half B\cdot  (n^2- n^2_0) \EE
so that, in units of $c=1$, the speed of sound is
\BE c^2_s=\frac{\partial P}{\partial
{\cal E}}\Big|_{\rm n=n_0} = \frac{d P/dn}{{d\cal E}  /dn} \Big|_{\rm n=n_0}=  \frac{B n_0 }{ A }\EE
By following Stevenson \cite{steveIJMPA}, one can envisage
two different regimes: a) the `empty vacuum' and b) the
`condensed vacuum'. Case a) corresponds to a very small
density of particles near the trivial empty state $n_0=0$ and is dominated by the rest mass term
${\cal E}(n) = m n + O(n^2)$. This limit has a vanishingly small
speed of sound \BE c^2_s=\frac{\partial P}{\partial
{\cal E}}\Big|_{\rm n=0}= 0  \EE The situation changes substantially in
the condensed vacuum where the effective potential $V_{\rm eff}
(\phi)$ gets its minimum at some $\phi=\pm v$. Then, due to
Eq.(\ref{neq}), the energy density ${\cal E}[n(\phi)]=V_{\rm eff}
(\phi)$ has its minimum at $n(\phi=\pm v)\equiv n_v$ where now $A=0$. Therefore, the speed of sound is formally infinite
\BE \label{infinite}c^2_s=\frac{\partial
P}{\partial {\cal E}}\Big|_{n=n_v}= \frac{Bn_v }{ A }\Big|_{\rm A=0}=  +\infty \EE
After that,
Stevenson's analysis \cite{steveIJMPA} goes
actually much farther, touching other aspects (as shock waves, post-hydrodynamic
approximations...) which go beyond the
scope of our paper. We thus address to ref.\cite{steveIJMPA} and
also to ref.\cite{steveCPT} for more details.

Of course,  the above elementary analysis cannot help to deduce a
(potentially) infinitely large $v_{QI}\to \infty $ from the
(potentially) infinitely large $c_s\to \infty $ of
Eq.(\ref{infinite}).  It just shows  that the physical vacuum medium
is incompressible or, better, that  it can support density
fluctuations whose wavelengths $\lambda$ will become larger and
larger in the $c_s \to \infty$ limit in order $c_s /\lambda$ can
remain finite. Thus, with sound waves of such long wavelengths, it
would be hard to produce the sharp wavefronts needed for
transmitting information.

Still, the argument indicates that the present view of the vacuum is
probably too narrow because, regardless of `messages', it is far
from obvious that $c$, the speed of light in this type of vacuum, is
a limiting speed. In this sense, it adds up to our discussion in
Sect.2, indicating that the idea of a Lorentz-invariant vacuum, and
therefore of the overall consistency with Einstein Special
Relativity without a preferred frame, is far from obvious.

And it also adds up to our analysis of the ether-drift experiments
in Sect.3, indicating that the standard null interpretation of the
data is, again, far from obvious once one starts to understand the
observed, irregular nature of the signal (compare e.g. the data in
Fig.\ref{Figcrossed} with our simulations in Fig.\ref{rotation} of the Appendix).
The required conceptual effort is modest, we believe, if compared
with the implications of the new perspective. In fact, by doing a
laser interferometry experiment, indoors inside a laboratory, from
the remarkable agreement of the experimental results
Eqs.(\ref{ravschiller2015}) and (\ref{ravnagelnature}) with the
theoretical predictions Eqs.(\ref{deltanuth}), (\ref{allanth}) and
(\ref{anysolid}), it is possible to perceive the motion of the solar
system, of our galaxy...within the background radiation.
Independently of the interpretation of relativity, this possibility
of perceiving reality in a global way (precisely in a completely
non-local way) is something fascinating. Clearly, this is implicit
in the idea of revealing our motion with respect to a privileged
reference system and fits well with the quantum view of correlations
over arbitrarily large distances. But this global vision of reality
perhaps goes beyond quantum correlations: it seems to have to do, in
a sense, with the quantum holographic principle \cite{zizzi1} that
all quantum information is "globalized". But perhaps it also has to
do with the vision of the internal observer, i.e. the observer
inside \cite{zizzi2} the quantum system, meaning that he is located
in a quantum space that is in a one-to-one relationship with the
quantum computational system under consideration.

\vskip 15 pt

\centerline{\Large \bf Appendix} \vskip 10 pt In this appendix, we
will summarize the simple stochastic model used in
refs.\cite{plus,plus2,book,universe} to compare with experiments.

To make explicit the time dependence of the signal let us first
re-write Eq.(\ref{bbasic2new}) as \begin{equation} \label{basic2}
     {{\Delta \bar{c}_\theta(t) } \over{c}}
    \sim
 \epsilon {{v^2(t) }\over{c^2}}\cos 2(\theta
-\theta_0(t)) \end{equation}  where $v(t)$ and $\theta_0(t)$
indicate respectively the instantaneous magnitude and direction of
the drift in the $(x,y)$ plane of the interferometer. This can also
be re-written as
\begin{equation} \label{basic3} {{\Delta \bar{c}_\theta(t) } \over{c}}\sim
2{S}(t)\sin 2\theta +
      2{C}(t)\cos 2\theta \end{equation} with \begin{equation} \label{amplitude10}
       2C(t)= \epsilon~ {{v^2_x(t)- v^2_y(t)  }
       \over{c^2}}~~~~~~~2S(t)=\epsilon ~{{2v_x(t)v_y(t)  }\over{c^2}}
\end{equation}  and $v_x(t)=v(t)\cos\theta_0(t)$, $v_y(t)=v(t)\sin\theta_0(t)$

As anticipated in Sect.3, the standard assumption to analyze the
data has always been based on the idea of regular modulations of the
signal associated with a cosmic Earth velocity. In general, this is
characterized by a magnitude $V$, a right ascension $\alpha$ and an
angular declination $\gamma$. These parameters can be considered
constant for short-time observations of a few days where there are
no appreciable changes due to the Earth orbital velocity around the
sun. In this framework, where the only time dependence is due to the
Earth rotation, the traditional identifications are $v(t)\equiv
\tilde v(t)$ and $\theta_0(t)\equiv\tilde\theta_0(t)$ where $\tilde
v(t)$ and $\tilde\theta_0(t)$ derive from the simple application of
spherical trigonometry \cite{nassau}
\begin{equation} \label{nassau1}
       \cos z(t)= \sin\gamma\sin \phi + \cos\gamma
       \cos\phi \cos(\tau-\alpha)
\end{equation} \begin{equation} \label{projection}
       \tilde {v}(t) =V \sin z(t)
\end{equation} \begin{equation} \label{nassau2}
    \tilde{v}_x(t) = \tilde{v}(t)\cos\tilde\theta_0(t)= V\left[ \sin\gamma\cos \phi -\cos\gamma
       \sin\phi \cos(\tau-\alpha)\right]
\end{equation} \begin{equation} \label{nassau3}
      \tilde{v}_y(t)= \tilde{v}(t)\sin\tilde\theta_0(t)= V\cos\gamma\sin(\tau-\alpha) \end{equation}
 Here $z=z(t)$ is the zenithal distance of
${\bf{V}}$, $\phi$ is the latitude of the laboratory,
$\tau=\omega_{\rm sid}t$ is the sidereal time of the observation in
degrees ($\omega_{\rm sid}\sim {{2\pi}\over{23^{h}56'}}$) and the
angle $\theta_0$ is counted conventionally from North through East
so that North is $\theta_0=0$ and East is $\theta_0=90^o$. With the
identifications $v(t)\equiv \tilde v(t)$ and
$\theta_0(t)\equiv\tilde\theta_0(t)$, one thus arrives to the simple
Fourier decomposition \begin{equation} \label{amorse1}
      S(t)\equiv {\tilde S}(t) =S_0+
      {S}_{s1}\sin \tau +{S}_{c1} \cos \tau
       + {S}_{s2}\sin(2\tau) +{S}_{c2} \cos(2\tau)
\end{equation}
\begin{eqnarray}
 \label{amorse2}
       C(t)\equiv {\tilde C}(t)=
       {C}_0 + {C}_{s1}\sin \tau +{C}_{c1} \cos \tau
       + {C}_{s2}\sin(2 \tau) +{C}_{c2} \cos(2 \tau)
\end{eqnarray}
where the  $C_k$ and $S_k$ Fourier coefficients depend on the three
parameters $(V,\alpha,\gamma)$ and are given explicitly in
refs.\cite{plus,book}.

Though, the identification of the instantaneous quantities $v_x(t)$
and $v_y(t)$ with their counterparts $\tilde{v}_x(t)$ and
$\tilde{v}_y(t)$ is not necessarily true. As anticipated in Sect.3,
one could consider the alternative situation where the velocity
field is a non-differentiable function and adopt some other
description, for instance a formulation in terms of random Fourier
series \cite{onsager,landau,fung}. In this other approach, the
parameters of the macroscopic motion are used to fix the typical
boundaries for a microscopic velocity field which has an intrinsic
non-deterministic nature.

The model adopted in refs.\cite{plus,plus2,book,universe}
corresponds to the simplest case of a turbulence which, at small
scales, appears homogeneous and isotropic. The analysis can then be
embodied in an effective space-time metric for light propagation
\begin{equation} \label{random} g^{\mu\nu}(t) \sim \eta^{\mu\nu} + 2
\epsilon v^\mu(t) v^\nu(t) \end{equation} where $v^\mu(t)$ is a
random 4-velocity field which describes the drift and whose
boundaries depend on a smooth field $\tilde{v}^\mu(t)$ determined by
the average Earth motion. By introducing the light 4-momentum $p_\mu$ and
replacing this metric in the relation $g^{\mu\nu}(t) p_\mu p_\nu=0$, one can determine
the one-way velocity $c_\gamma(\theta)=p_0(\theta)/|{\bf p}|$ and then the two-way
combination through Eq.(\ref{twoway0}).

For homogeneous turbulence a series representation, suitable for
numerical simulations of a discrete signal, can be expressed in the
form
\begin{equation} \label{vx} v_x(t_k)= \sum^{\infty}_{n=1}\left[
       x_n(1)\cos \omega_n t_k + x_n(2)\sin \omega_n t_k \right] \end{equation}
\begin{equation} \label{vy} v_y(t_k)= \sum^{\infty}_{n=1}\left[
       y_n(1)\cos \omega_n t_k + y_n(2)\sin \omega_n t_k \right] \end{equation}
Here $\omega_n=2n\pi/T$ and T is the common period of all Fourier
components. Furthermore, $t_k= (k-1) \Delta t$, with $k=1, 2...$,
and $\Delta t$ is the sampling time. Finally, $x_n(i=1,2)$ and
$y_n(i=1,2)$ are random variables with the dimension of a velocity
and vanishing mean. In our simulations, the value $T=T_{\rm day}$=
24 hours and a sampling step $\Delta t=$ 1 second were adopted.
However, the results would remain unchanged by any rescaling $T\to s
T$ and $\Delta t\to s \Delta t$.

In general, we can denote by $[-d_x(t),d_x(t)]$ the range for
$x_n(i=1,2)$ and by $[-d_y(t),d_y(t)]$ the corresponding range for
$y_n(i=1,2)$. Statistical isotropy would require to impose $d_x(t)=
d_y(t)$. However, to illustrate the more general case, we will first
consider $d_x(t) \neq d_y(t)$.

If we assume that the random values of $x_n(i=1,2)$ and $y_n(i=1,2)$
are chosen with uniform probability, the only non-vanishing
(quadratic) statistical averages are
\begin{equation} \label{quadratic} \langle x^2_n(i=1,2)\rangle_{\rm
stat}={{d^2_x(t) }\over{3 ~n^{2\eta}}}~~~~~~~~~~~\langle
y^2_n(i=1,2)\rangle_{\rm stat}={{d^2_y(t) }\over{3 ~n^{2\eta}}}
\end{equation} Here, the exponent $\eta$ ensures
finite statistical averages  $\langle v^2_x(t)\rangle_{\rm stat}$
and $\langle v^2_y(t)\rangle_{\rm stat}$ for an arbitrarily large
number of Fourier components. In our simulations, between the two
possible alternatives $\eta=5/6$ and $\eta=1$ of ref.\cite{fung}, we
have chosen $\eta=1$ that corresponds to the Lagrangian picture in
which the point where the fluid velocity is measured is a wandering
material point in the fluid.

Finally, the connection with the Earth cosmic motion is obtained by
identifying $d_x(t)=\tilde v_x(t)$ and $d_y(t)=\tilde v_y(t)$ as
given in Eqs. (\ref{nassau1})$-$(\ref{nassau3}). If, however, we
require statistical isotropy, the relation
\begin{equation} \label{correct} \tilde{v}^2_x(t) +
\tilde{v}^2_y(t)=\tilde{v}^2(t)\end{equation} requires the
identification
\begin{equation} \label{isot} d_x(t)=d_y(t)={{ \tilde{v}(t)}\over{\sqrt{2} }} \end{equation}
 For such isotropic model, by combining
Eqs.(\ref{vx})$-$(\ref{isot}) and in the limit of an infinite
statistics, one gets
\begin{eqnarray}
\label{vanishing} \langle v^2_x(t)\rangle_{\rm stat}=\langle
v^2_y(t)\rangle_{\rm stat}={{\tilde{v}^2(t)}\over{2}}~{{1}\over{3}}
\sum^{\infty}_{n=1} {{1}\over{n^2}}= {{\tilde{v}^2(t)}\over{2}}~
{{\pi^2}\over{18}}\nonumber \\ \langle v_x(t)v_y(t)\rangle_{\rm
stat}=0
\end{eqnarray}
and  vanishing statistical averages
\begin{equation} \label{vanishing2}\langle C(t)\rangle_ {\rm
stat}=0~~~~~~~~~~~~~~~~~~\langle S(t)\rangle_ {\rm stat}=0
\end{equation} at {\it any} time $t$, see Eqs.(\ref{amplitude10}).
Therefore, by construction, this model gives a definite non-zero
signal but, if the same signal were fitted with Eqs.(\ref{amorse1})
and (\ref{amorse2}), it would also give average values $(C_k)^{\rm
avg}=0$, $(S_k)^{\rm avg}=0$ for the Fourier coefficients.

\begin{figure}[h]
\begin{center}
\includegraphics[scale=0.26]{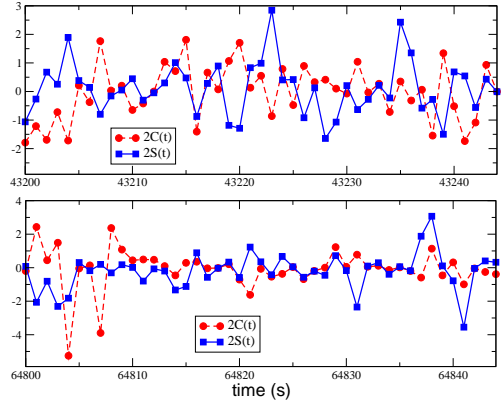}
\end{center}
\caption{\it For $\epsilon_v$ as in Eq.(\ref{refractive2}) and
$\chi=2$, we report in units $10^{-15}$ two typical sets of 45
seconds for the two functions $2C(t)$ and $2S(t)$ of
Eq.(\ref{basic3}). The two sets belong to the same random sequence
and refer to two sidereal times that differ by 6 hours. The
boundaries of the stochastic velocity components Eqs.(\ref{vx}) and
(\ref{vy}) are controlled by $(V,\alpha,\gamma)_{\rm CMB}$ through
Eqs.(\ref{projection}) and (\ref{isot}). For a laser frequency of
$2.8\cdot 10^{14}$ Hz, the range $\pm 3.5\cdot 10^{-15}$ corresponds
to a typical frequency shift $\Delta \nu$ in the range $\pm 1$ Hz,
as in our Fig.\ref{Figcrossed}.} \label{rotation}
\end{figure}

To understand how radical is the modification produced by
Eqs.(\ref{vanishing2}), we recall the traditional procedure adopted
in the classical experiments. One was measuring the fringe shifts at
some given sidereal time on consecutive days so that changes of the
orbital velocity were negligible. Then, see Eqs.(\ref{newintro1})
and (\ref{basic3}), the measured shifts at the various angle
$\theta$ were averaged
\begin{equation} \label{averagefringe}\langle{{\Delta
\lambda(\theta;t)}\over{\lambda}}\rangle_ {\rm stat}=
{{2D}\over{\lambda}} \left[2\sin 2\theta~\langle S(t)\rangle_ {\rm
stat} + 2\cos 2\theta~\langle C(t)\rangle_ {\rm stat} \right]
\end{equation} and finally these average values were compared with
models for the Earth cosmic motion.

However if the signal is so irregular that, by increasing the number
of measurements, $\langle C(t)\rangle_ {\rm stat} \to 0$ and
$\langle S(t)\rangle_ {\rm stat} \to 0$ the averages
Eq.(\ref{averagefringe}) would have no meaning. In fact, these
averages would be non vanishing just because the statistics is
finite. In particular, the direction $\theta_0(t)$ of the drift
(defined by the relation $\tan2\theta_0(t)= S(t)/C(t)$) would vary
randomly with no definite limit.

This is why one should concentrate the analysis on the 2nd-harmonic
amplitudes
\begin{equation} \label{AA} A_2(t)={{2D}\over{\lambda}}~
2\sqrt{S^2(t) + C^2(t)} \sim {{2 D }\over{\lambda}}~ \epsilon {{
v^2_x(t)+v^2_y(t)} \over{c^2 }}
\end{equation}
which are positive-definite and remain non-zero under the averaging
procedure. Moreover, these are rotational-invariant quantities and
their statistical average \BE \label{amplitude10001finalapp1}
\langle A_2(t)\rangle_{\rm stat} \sim  {{2
D }\over{\lambda}} \cdot {{\pi^2 } \over{18 }}\cdot \epsilon\cdot \frac{V^2 \sin^2 z(t)}{c^2}
\EE would remain unchanged in the isotropic model Eq.(\ref{isot}) or
with the alternative choice $d_x(t)\equiv \tilde v_x(t)$ and
$d_y(t)\equiv \tilde v_y(t)$.

Analogous considerations hold for the modern experiments where
$\frac{\Delta \bar{c}_\theta(t)}{c}$ is extracted from the frequency
shift of two optical resonators. Again, the $C(t)$ and $S(t)$
obtained, through Eq.(\ref{basic3}), from the very irregular signal
(see e.g. Fig.\ref{Figcrossed}), are compared with the slowly
varying parameterizations Eqs.(\ref{amorse1}) and (\ref{amorse2}).
No surprise that the average values $(C_k)^{\rm avg}=0$, $(S_k)^{\rm
avg}=0$ of the resulting Fourier coefficients become smaller and
smaller by simply increasing the number of observations. To 
appreciate the change of perspective in our stochastic model,
compare with a simulation of the $C(t)$ and $S(t)$ in
Fig.\ref{rotation}.




\begin{thebibliography}{99}

\bibitem{weinberg} S. Weinberg, {\it The Trouble with
Quantum Mechanics}, The New York Review of Books, January 19, 2017.
\bibitem{BFS}
Ph. Blanchard, J. Fr\"ohlich, and B. Schubnel, {\it A ``garden of
forking paths'' -  The quantum mechanics of histories of events}
Nucl. Phys. B \textbf{912} (2016) 463; arXiv:1603.09664 [quant-ph].
\bibitem{EPR}
A. Einstein, B. Podolski and N. Rosen, {\it Can quantum-mechanical
description of physical reality be considered complete?}, Phys.
Rev.\textbf{47} (1935) 777.
\bibitem{Bell}
Full reference to all papers by J. S. Bell, can be found in the
Volume Collection, {\it Speakable and Unspeakable in Quantum
Mechanics}, 2nd edition, Cambridge University Press, 2004.
\bibitem{Stapp}
H. P. Stapp, {\it A Bell-type theorem without hidden variables}, Am.
Journ. of Physics, \textbf{72} (2004) 30.
\bibitem{Shimony}
A. Shimony, {\it An Analysis of Stapp's ``A Bell-type theorem
without hidden variables''}, arXiv:quant-ph/0404121.
\bibitem{Bricmont0}
J. Bricmont, {\it What Did Bell Really Prove?}, in Quantum
NonLocality And Reality, 50 Years of Bell's theorem, M. Bell and S.
Gao. Eds. Cambridge Univ. Press 2016, p. 49.
\bibitem{Maudlin}
T. Maudlin, {\it Quantum Non-Locality and Relativity}, Blackwell,
Cambridge, 2011.
\bibitem{Dirac}
P. A. M. Dirac, {\it Development of the Physicist's Conception of
Nature}, in The Physicist's Conception of Nature, J. Mehra Ed.,
Reidel, Boston 1973.
\bibitem{Weinbergsound}
S. Weinberg, {\it Gravitation and Cosmology}, John Wiley and Sons,
Inc., 1972, pag. 52.
\bibitem{undivided}
D. Bohm and B. Hiley, {\it The Undivided Universe}, Routledge,
London 1993.
\bibitem{Hardy}
L. Hardy, {\it Quantum mechanics, local realistic theories, and
Lorentz-invariant realistic theories}, Phys. Rev. Lett. {\bf 68}
(1992) 2981.
\bibitem{Caban}
P. Caban and J. Rembielinski, {\it Lorentz-covariant quantum
mechanics and preferred frame}, Phys. Rev. A\textbf{59} (1999) 4187.
\bibitem{Sonego}
S. Liberati, S. Sonego and M. Visser, {\it Faster-than-c signals,
special relativity, and causality}, Ann. Phys.\textbf{298} (2002)
167.
\bibitem{Eberhard1}
P. H. Eberhard, {\it Bell's Theorem and the Different Concept of
Locality}, Nuovo Cim.B \textbf{46} (1978) 392.
\bibitem{Eberhard2}
P. H. Eberhard, {\it A realistic model for Quantum Theory with a
locality property}, in Quantum Theories and Pictures of Reality, W.
Schommers Ed., Springer Verlag, Berlin (1989), p.169.
\bibitem{garisto}
R. Garisto, {\it What is the speed of quantum information?},
arXiv:quant-ph/0212078.
\bibitem{yoon}
M. Yoon and D. Huterer, {\it Kinematic Dipole Detection With Galaxy
Surveys: Forecasts And Requirements}, Astrophys. J. Lett. {\bf 813}
(2015) L18.
\bibitem{scarani}
V. Scarani, W. Tittel, H. Zbinden and N. Gisin, {\it The speed of
quantum information and the preferred frame: analysis of
experimental data}, Phys. Lett. A \textbf{276} (2000) 1.
\bibitem{cocciaro}
B. Cocciaro, S. Faetti and L. Fronzoni, {\it A lower bound for the
velocity of quantum communication in the preferred frame}, Phys.
Lett. A\textbf{375} (2011) 379.
\bibitem{salart}
D. Salart, A. Baas, C. Branciard, N. Gisin and H. Zbinden, {\it
Testing spooky action at a distance}, Nature \textbf{454} (2008)
861.
\bibitem{bancal}
J.-D. Bancal, S. Pironio, A. Acin, Y.-C. Liang, V. Scarani, and N.
Gisin, {\it Quantum nonlocality based on finite-speed causal
influences leads to superluminal signaling}, Nature Physics
\textbf{8} (2012) 867.
\bibitem{cocciaro2}
B. Cocciaro, S. Faetti and L. Fronzoni, {\it  Fast measurements of entanglement over a
kilometric distance to test superluminal models of
Quantum Mechanics: final results }, J. Phys. Conf. Ser.
\textbf{1275} (2019)  012035.
\bibitem{maiani}
L. Maiani and M. Testa, {\it Causality in quantum field theory},
Phys. Lett. B{\bf 356} (1995) 319.
\bibitem{Bricmont}
J. Bricmont, {\it Making Sense of Quantum Mechanics}, Springer
Int. Publ. 2016.
\bibitem{thooft} G. 't Hooft, {\it Search of the
Ultimate Building Blocks}, Cambridge Univ. Press 1997, p.70.
\bibitem{mech}
M.~Consoli, P.M. Stevenson, {\it Physical mechanisms generating
spontaneous symmetry breaking and a hierarchy of scales}, Int. J.
Mod. Phys. A \textbf{15}(2000) 133, hep-ph/9905427.
\bibitem{Kostro}
L. Kostro, {\it Einstein and the Ether}, Italian translation, Ed.
Dedalo, Bari 2001.
\bibitem{Chiao}
R. Y. Chiao, {\it Conceptual tensions between quantum mechanics and
general relativity: Are there experimental consequences?}, in
``Science and Ultimate Reality: From Quantum to Cosmos'', honoring
John Wheeler's 90th birthday. J. D. Barrow, P. C. W. Davies, and C.
L. Harper eds. Cambridge University Press (2003);
arXiv:gr-qc/0303100.
\bibitem{jauch}
J. M. Jauch and K. M. Watson, {\it Phenomenological
Quantum-Electrodynamics}, Phys. Rev. {\bf 74}, 950 (1948).
\bibitem{plus}
M. Consoli, C. Matheson and A. Pluchino, {\it The classical
ether-drift experiments: a modern re-interpretation} Eur. Phys. J.
Plus, {\bf 128} (2013) 71.
\bibitem{plus2}
M. Consoli and A. Pluchino, {\it Cosmic Microwave Background and the
issue of a fundamental preferred  frame}, Eur. Phys. Jour. Plus {\bf
133} (2018) 295.
\bibitem{book}
M. Consoli and A. Pluchino, {\it Michelson-Morley Experiments: an
Enigma for Physics and the History of Science}, World Scientific
2019, ISBN 978-981-3278-18-9.
\bibitem{universe}
M. Consoli and A. Pluchino, {\it CMB, preferred reference system and
dragging of light in the earth's frame}, Universe {\bf 7} (2021)
311;arXiv:2109.03047 [physics.gen-ph].
\bibitem{epl}
M. Consoli, A. Pluchino and A. Rapisarda, {\it Cosmic Background
Radiation and `ether-drift' experiments}, Europhysics Lett. {\bf
113} (2016) 19001.
\bibitem{CW}
S.R. Coleman, E.J. Weinberg, {\it Radiative Corrections as the
Origin of Spontaneous Symmetry Breaking}, Phys. Rev. D\textbf{7}
(1973) 1888.
\bibitem{lattice1}
P.H. Lundow, K.~Markstr{\"o}m, {\it Critical behavior of the Ising
model on the four-dimensional cubic lattice}, Physical Review E
\textbf{80} (2009) 031104.
\bibitem{lattice2}
P.H. Lundow, K.~Markstr{\"o}m, {\it Non-vanishing boundary effects
and quasi-first order phase transitions in high dimensional Ising
models}, Nucl. Phys. B\textbf{845} (2011) 120.
\bibitem{lattice3}
S. Akiyama, Y. Kuramashi, T. Yamashita, Y. Yoshimura, {\it Phase
transition of four-dimensional Ising model with higher-order tensor
renormalization group},  Phys. Rev. D \textbf{100} (2019) 054510.
\bibitem{cpt}
See, for instance, R. F. Streater and  A. S. Wightman, {\it PCT,
Spin and Statistics, and all that}, W. A. Benjamin, New York 1964.
\bibitem{segal}
I.E. Segal, {\it Is the Physical Vacuum Really Lorentz-Invariant?},
in Differential Geometry, Group Representations, and Quantization,
J.D. Hennig, W.L\"ucke, J. Tolar, Eds.  Lecture Notes
in Physics Vol. 379, Springer 1991.
\bibitem{Stefanovich}
E. V. Stefanovich, {\it Is Minkowski Space-Time Compatible with
Quantum Mechanics?}, Found. Phys.\textbf{32} (2002) 673.
\bibitem{Glazek}
S. D. Glazek and T. Maslowski, {\it Renormalized Poincar\'e algebra
for effective particles in quantum field theory} Phys. Rev.
D\textbf{65} (2002) 065011.
\bibitem{ciancitto}
M. Consoli and A. Ciancitto, {\it Indications of the occurrence of
spontaneous symmetry breaking in massless $\lambda \Phi^4$ theory},
 Nucl. Phys. B{\bf 254}, 653 (1985).
\bibitem{weinbergQFT}
S. Weinberg, {\it The Quantum Theory of Fields}, Cambridge
University Press, Vol.II, pp. 163-167.
\bibitem{epjc}
M. Consoli and E. Costanzo, {\it Is the physical vacuum a preferred
frame?}, Eur. Phys. Journ. {\bf C54} (2008) 285.
\bibitem{dedicated}
M. Consoli and E. Costanzo, {\it Precision tests with a new class of
dedicated ether-drift experiments}, Eur. Phys. Journ. {\bf C55}
(2008) 469.
\bibitem{foop}
M. Consoli,{\it Probing the vacuum of particle physics with precise
laser interferometry}, Found. of Phys. {\bf 45}, 22 (2015).
\bibitem{mpla11}
M. Consoli,{\it On the low-energy spectrum of spontaneously broken
$\Phi^4$ theories }, Mod. Phys. Lett. A {\bf 26} (2011) 531.
\bibitem{zeldovich}
Y. B. Zeldovich, {\it  The Cosmological constant and the theory of elementary particles},
Sov. Phys. Usp. {\bf 11}, 381 (1968).
\bibitem{weinbergreview}
S. Weinberg, {\it  The cosmological constant problem},
Rev. Mod. Phys. {\bf 61}, 1 (1989).
\bibitem{smoot}
G. F. Smoot, {\it Cosmic microwave background radiation
anisotropies: Their discovery and utilization}, Nobel Lecture, Rev.
Mod. Phys. {\bf 79}, 1349 (2007).
\bibitem{ungar}
A. Ungar,{\it  The relativistic composite-velocity reciprocity
principle}, Found. of Phys. {\bf 30}, 331 (2000).
\bibitem{costella}
J. P. Costella et al.,{\it  The Thomas rotation}, Am. J. Phys. {\bf
69}, 837 (2001).
\bibitem{kanevisser}
K. O' Donnell and M. Visser,{\it Elementary analysis of the special
relativistic combination of velocities, Wigner rotation, and Thomas
precession}, Eur. J. Phys. {\bf 32}, 1033 (2011).
\bibitem{nagelnature}
M. Nagel  et al., {\it Direct terrestrial test of Lorentz symmetry
in electrodynamics to $10^{-18}$}, Nature Comm.{\bf 6}, 8174 (2015).
\bibitem{mm}
A. A. Michelson and E. W. Morley, {\it On the Relative Motion of the
Earth and the Luminiferous Ether}, Am. J. Sci. {\bf 34}, 333 (1887).
\bibitem{miller}
D. C. Miller, {\it The Ether-Drift Experiment and the Determination
of the Absolute Motion of the Earth}, Rev. Mod. Phys. {\bf 5}, 203
(1933).
\bibitem{kenconference}
A. A. Michelson, et al., {\it Conference on the Ether-Drift
Experiments}, Ap. J. {\bf 68} (1928) p. 341-402.
\bibitem{illingworth}
K. K. Illingworth, {\it A Repetition of the Michelson-Morley
Experiment Using Kennedy's Refinement}, Phys. Rev. {\bf 30}, 692
(1927).
\bibitem{tomaschek1}
R. Tomaschek, {\it About the Michelson experiment with fixed star
light}, Astron. Nachrichten, {\bf 219}, 301 (1923), English
translation.
\bibitem{piccard3}
A. Piccard and E. Stahel, {\it REALIZATION OF THE EXPERIMENT OF
MICHELSON IN BALLOON AND ON DRY LAND}, Journ. de Physique  et Le
Radium {\bf IX} (1928) No.2.
\bibitem{mpp}
A. A. Michelson, F. G. Pease and F. Pearson,{\it Repetition of the
Michelson-Morley Experiment}, Nature, {\bf 123}, 88 (1929).
\bibitem{mpp2}
A. A. Michelson, F. G. Pease and F. Pearson, {\it Repetition of the
Michelson-Morley experiment}, J. Opt. Soc. Am. {\bf 18}, 181 (1929).
\bibitem{pease}
F. G. Pease, {\it Ether-Drift Data}, Publ. of the Astr. Soc. of the
Pacific, {\bf XLII}, 197 (1930).
\bibitem{joos}
G. Joos, {\it Die Jenaer Wiederholung des Michelsonversuchs}, Ann.
d. Physik {\bf 7}, 385 (1930).
\bibitem{kennedy}
R. J. Kennedy,{\it Simplified theory of the Michelson-Morley
experiment}, Phys. Rev. {\bf 47}, 965 (1935).
\bibitem{maxwell}
J. C. Maxwell, {\it Ether}, Encyclopaedia Britannica, 9th Edition, 1878.
\bibitem{chaos}
M. Consoli, A. Pluchino and A. Rapisarda,{\it Basic randomness of
nature and ether-drift experiments}, Chaos, Solitons and Fractals
{\bf 44}, 1089 (2011).
\bibitem{physica}
M. Consoli, A. Pluchino, A. Rapisarda and S. Tudisco, {\it The
vacuum as a form of turbulent fluid: motivations, experiments,
implications}, Physica {\bf A394}, 61 (2014).
\bibitem{whittaker}
E. T. Whittaker,{\it A History of the Theories of Aether and
Electricity}, Dover Publ., New York 1989.
\bibitem{troshkin}
O. V. Troshkin, {\it  On wave properties of an incompressible turbulent fluid},
Physica A {\bf 168} (1990) 881.
\bibitem{puthoff}
H. E. Puthoff,{\it Linearized turbulent flow as an analog model for
linearized General Relativity}, arXiv:0808.3401 [physics.gen-ph].
\bibitem{tsankov}
T. D. Tsankov,{\it  Classical Electrodynamics and the Turbulent Aether
Hypothesis}, Preprint February 2009, unpublished.
\bibitem{saul}
L. A. Saul, {\it Spin Waves as Metric in a Kinetic Space-Time},
Phys. Lett. {\bf A 314} (2003) 472.
\bibitem{pla12}
M. Consoli, {\it A kinetic basis for space-time symmetries}, 
Phys. Lett. {\bf A376} (2012) 3377.
\bibitem{nelson}
E. Nelson, {\it A derivation of the Schr\"odinger Equation from
Newtonian Mechanics}, Phys. Rev. {\bf 150} (1966) 1079.
\bibitem{kleinert}
P. Jizba and H. Kleinert,
{\it Superstatistics approach to path integral for a relativistic particle}
Phys. Rev. D {\bf 82} (2010) 085016.
\bibitem{jizba}
P. Jizba, F. Scardigli, {\it Special Relativity induced by
Granular Space}, Eur. Phys. J.  C{\bf 73} (2013) 2491.
\bibitem{feybook}
R. P. Feynman, R. B. Leighton and M. Sands, {\it The Feynman
Lectures on Physics}, Addison Wesley Publ. Co. 1963.
\bibitem{onsager}
L. Onsager,  N. Cimento, {\it Statistical hydrodynamics}, Suppl.
{\bf 6}, 279 (1949).
\bibitem{eyink}
G. L. Eyink and K. R. Sreenivasan, {\it Onsager and the theory of
hydrodynamic turbulence}, Rev. Mod. Phys. {\bf 78}, 87 (2006).
\bibitem{nassau}
J. J. Nassau and P. M. Morse,{\it A Study of Solar Motion by
Harmonic Analysis}, Ap. J. {\bf 65}, 73 (1927).
\bibitem{landau}
L. D. Landau and E. M. Lifshitz,{\it  Fluid Mechanics}, Pergamon
Press 1959.
\bibitem{fung}
J. C. H. Fung et al., {\it Kinematic simulation of homogeneous
turbulence by unsteady random Fourier modes}, J. Fluid Mech. {\bf
236}, 281 (1992).
\bibitem{shankland}
R. S. Shankland et al., {\it New Analysis of the Interferometer
Observations of Dayton C. Miller}, Rev. Mod. Phys.{\bf 27}, 167
(1955).
\bibitem{swensonbook}
L. S. Swenson Jr., {\it The Ethereal Aether, A History of the
Michelson-Morley-Miller Aether-Drift Experiments}, 1880-1930.
University of Texas Press, Austin 1972.
\bibitem{loyd2}
Loyd S. Swenson Jr.,{\it The Michelson-Morley-Miller Experiments
before and after 1905}, Journ. for the History of Astronomy, {\bf
1}, 56 (1970).
\bibitem{fox}
J. Shamir and R. Fox, {\it A New Experimental Test of Special
Relativity}, N. Cim. {\bf 62B}, 258 (1969).
\bibitem{joos2}
G. Joos, {\it Note on the Repetition of the Michelson-Morley
Experiment}, Phys. Rev. {\bf 45}, 114 (1934).
\bibitem{miller34}
D. C. Miller, {\it Comments on Dr. Georg Joos's Criticism of the
Ether-Drift Experiment}, Phys. Rev. {\bf 45} (1934) 114.
\bibitem{farkas}
E. R. Farkas and W. W. Webb, {\it Precise and millidegree stable
control for fluorescense imaging}, Rev. Scient. Instr. 81, 093704
(2010).
\bibitem{zhaoa}
Y. Zhao, D. L. Trumper, R. K. Heilmann, M. L. Schattenburg, {\it
Optimizatiom and temperature mapping of an ultra-high thermal
stability enviromental enclosure}, Precision Engin. 34, 164 (2010).
\bibitem{trusov}
I. P. Prikhodko, A. A. Trusov, A. M. Shkel, {\it Compensation of
drifts in high-Q MEMS gyroscopes using temperature self-sensing},
Sensors and Actuators A 201, 517 (2013).
\bibitem{mueller2003}
H. M\"uller, et al. , {\it Modern Michelson-Morley Experiment using
Cryogenic Optical Resonators}, Phys. Rev. Lett. {\bf 91}, 020401
(2003).
\bibitem{crossed}
Ch. Eisele, M. Okhapkin, A. Nevsky, S. Schiller, {\it  A crossed optical cavities apparatus for a
precision test of the isotropy of light propagation}, Opt. Comm.
{\bf 281}, 1189 (2008).
\bibitem{newberlin}
S. Herrmann, et al., {\it Rotating optical cavity experiment testing
Lorentz invariance at the $10^{-17}$ level}, Phys.Rev. D {\bf 80},
10511 (2009).
\bibitem{newschiller}
Ch. Eisele, A. Newsky  and S. Schiller, {\it Laboratory Test of the
Isotropy of Light Propagation at the $10^{-17}$ Level}, Phys. Rev.
Lett. {\bf 103}, 090401 (2009).
\bibitem{cpt2013}
M. Nagel et al.,{\it  Ultra-stable Cryogenic Optical Resonators For
Tests Of Fundamental Physics}, arXiv:1308.5582[physics.optics].
\bibitem{schiller2015}
Q. Chen, E. Magoulakis, and S. Schiller,{\it High-sensitivity
crossed-resonator laser apparatus for improved tests of Lorentz
invariance and of space-time fluctuations}, Phys. Rev. D {\bf 93 },
022003 (2016).
\bibitem{numata}
K. Numata, A, Kemery and J. Camp, {\it Thermal-Noise Limit in the
Frequency Stabilization of Lasers with Rigid Cavities}, Phys. Rev.
Lett. {\bf 93}, 250602 (2004).
\bibitem{applied}
H. M\"uller, et al., {\it Precision test of the isotropy of light propagation}, Appl. Phys. B {\bf 77}, 719 (2003).
\bibitem{gerg}
M. Consoli and L. Pappalardo, {\it Emergent gravity and ether-drift
experiments}, Gen. Rel. and Grav. {\bf 42}, 2585 (2010).
\bibitem{broekaert}
J. Broekaert, {\it A Spatially-VSL Gravity Model with 1-PN limit of
GRT}, Found. of Phys. {\bf 38}, 409 (2008).
\bibitem{eddington}
A. S. Eddington,{\it Space, Time and Gravitation}, Cambridge
University Press, 1920.
\bibitem{landaufield}
L. D. Landau and E. M. Lifshitz,{\it The Classical Theory of
Fields}, Pergamon Press, 1971, p.257.
\bibitem{optis}
C. L\"ammerzahl et al., {\it  OPTIS: a satellite-based test of
special and general relativity},Class. Quantum Gravity {\bf 18},
2499 (2001).
\bibitem{KP}
D. A. Kirzhnits and V. L. Polyachenko, {\it ON THE POSSIBILITY OF
MACROSCOPIC MANIFESTATIONS OF VIOLATION OF MICROSCOPIC CAUSALITY},
Sov. Phys. JETP {\bf 19}, 514 (1964).
\bibitem{BR1}
S. A. Bludman and M. A. Ruderman, {\it Possibility of the Speed of
Sound Exceeding the Speed of Light in Ultradense Matter}, Phys. Rev.
{\bf 170}, 1176 (1968).
\bibitem{R} M. Ruderman, {\it Causes of Sound
Faster than Light in Classical Models of Ultradense Matter}, Phys.
Rev. {\bf 172}, 1286 (1968).
\bibitem{BR2}
S. A. Bludman and M. A. Ruderman, {\it Noncausality and Instability
in Ultradense Matter}, Phys. Rev. D{\bf 1}, 3243 (1970).
\bibitem{kp}
B. D. Keister and W. N. Polyzou, {\it Causality in dense matter},
Phys. Rev. C{\bf 54}, 2023 (1996).
\bibitem{paulsound}
P. M. Stevenson, {\it How do sound waves
in a Bose-Einstein condensate move so fast?}, Phys. Rev. A{\bf 68} (2003) 055601.
\bibitem{steveIJMPA}
P. M. Stevenson, {\it HYDRODYNAMICS OF THE VACUUM}, Int. J. Mod.
Phys. A {\bf 21}, 2877 (2006).
\bibitem{steveCPT}
P. M. Stevenson, {\it Are There Pressure Waves in the Vacuum?}, in
Proceedings of the Second Meeting on CPT and Lorentz Symmetry, V. A.
Kostelecky Ed., World Scientific, Singapore, 2002;
arXiv:hep-ph/0109204.
\bibitem{zizzi1}
P. Zizzi, {\it Quantum Holography from Fermion Fields}, Quantum Rep.
2021, 3(3), 576-591; https://doi.org/10.3390/quantum3030037.
\bibitem{zizzi2}
P. Zizzi, {\it Consciousness and logic in a quantum computing
universe}, in "The emerging physics of consciousness", Jack
Tuszynski Ed., Springer, 2006.
\end{thebibliography}
\end{document}